\documentclass{aa}  
\usepackage{graphicx}
\usepackage{multirow}
\newcommand{\bm}[1]{{\mbox{\boldmath $#1$}}}

\usepackage{color}
\usepackage{here}

\newcommand{\ii}{{{\rm i}}}

\usepackage[colorlinks=true,citecolor=blue]{hyperref}
\usepackage{arydshln}
\usepackage{natbib}

\usepackage{comment}

\usepackage{booktabs}

\usepackage{adjustbox}
\usepackage{academicons}
\usepackage{orcidlink}

\usepackage{ulem}

\usepackage{scalerel}
\usepackage{stackengine,wasysym}

\usepackage{txfonts}
\begin{document} 
   \title{
   Impacts of small-scale dynamo on rotating columnar \\ 
   convection in stellar convection zones
   }

   \author{Yuto Bekki \orcidlink{0000-0002-5990-013X} 
          }

   \institute{Max-Planck-Institut f{\"u}r Sonnensystemforschung,
              Justus-von-Liebig-Weg 3, 37077 G{\"o}ttingen, Germany\\
            \email{\href{mailto:bekki@mps.mpg.de}{bekki@mps.mpg.de}}
             }

   \date{Received <-->; accepted <-->}


  \abstract
{
Understanding the complex interactions between convection, magnetic fields, and rotation is key to modeling the internal dynamics of the Sun and stars. Under rotational influence, compressible convection forms prograde-propagating convective columns near the equator. The interaction between such rotating columnar convection and the small-scale dynamo (SSD) remains largely unexplored.
} 
{
We investigate the influence of the SSD on the properties of rotating convection in the equatorial regions of solar and stellar convection zones.
}
{
A series of rotating compressible magnetoconvection simulations is performed using a local $f$-plane box model at the equator. The flux-based Coriolis number Co$_*$ is varied systematically. To isolate the effects of the SSD, we compare results from hydrodynamic (HD) and magnetohydrodynamic (MHD) simulations.
}
{
The SSD affects both convective heat and angular momentum transport. In MHD cases, convective velocity decreases more rapidly with increasing Co$_*$ than in HD cases. This reduction is compensated by enhanced entropy fluctuations, maintaining overall heat transport efficiency. Furthermore, a weakly subadiabatic layer is maintained near the base of the convection zone even under strong rotational influence when the SSD is present. These behaviors reflect a change in the dominant force balance: the SSD introduces a magnetostrophic balance at small scales, while geostrophic balance persists at larger scales. The inclusion of the SSD also reduces the dominant horizontal scale of columnar convective modes by enhancing the effective rotational influence. Regarding angular momentum transport, the SSD generates Maxwell stresses that counteract the Reynolds stresses, thereby quenching the generation of mean shear flows.
}
{
Small-scale magnetic fields interact nonlinearly with columnar convection and induce substantial modifications in the dynamics of rotating convection. These effects should be accounted for in models of solar and stellar convection.
}


   \maketitle
%


\section{Introduction} \label{sec:intro}

The Sun's convection zone is differentially rotating \citep[e.g.,][]{thompson1996}. This differential rotation is believed to be generated by the nonlinear interaction between rotation and magneto-convection. The solar-like differential rotation, with a faster equator and slower poles, has been repeatedly produced in three-dimensional (3D) numerical simulations of rotating convection in spherical shells \citep[e.g.,][]{brun2002,miesch2008,kapyla2011,guerrero2013,hotta2014b}. In these numerical models, a key parameter controlling the differential rotation regime is the Rossby number $\mathrm{Ro}=v/(2\Omega \ell)$ or Coriolis number $\mathrm{Co}=\mathrm{Ro}^{-1}$, where $v$ and $\ell$ are the typical velocity and length scales of convection, and $\Omega$ is the rotation rate.
In a high-Ro (or equivalently low-Co) regime, the differential rotation tends to become anti-solar with faster poles and a slower equator. On the other hand, in a low-Ro (high-Co) regime, the differential rotation tends to become solar-like \citep[e.g.,][]{gastine2013,mabuchi2015,featherstone2015,viviani2018,brun2022}. The transition between solar-like and anti-solar rotational profiles occurs at around $\mathrm{Ro} \approx 1$ \citep[][]{gastine2013,mabuchi2015,viviani2018}. The problem of numerical modeling of the Sun's global-scale convection is that the typical Ro of the Sun is likely very close to this transition \citep[e.g.,][]{strugarek2017,vasil2021}, and therefore, numerical results are highly dependent on subtle changes in the model parameters. In particular, the anti-solar differential rotation is preferentially obtained in recent highly turbulent models when the solar values of rotation rate $\Omega_{\odot}$ and luminosity $L_{\odot}$ are used. This is, in fact, a striking aspect of the Sun's convective conundrum \citep[][]{hanasoge2012,omara2016,hotta2023_review,kapyla2023_review}.

A physical origin of the solar-like differential rotation in the low-Ro regime can be explained as follows. Under strong rotational influence, convection tends to be aligned with the rotational axis and takes the form of columnar rolls outside the tangential cylinder \citep[][]{busse1970}. They are called Busse columns or banana cells and can be readily seen in simulations as north–south-aligned lanes of downflows across the equator \citep[][]{busse1970,gilman1986,miesch2000}. These banana cells have an associated Reynolds stress which transports the angular momentum equatorward and cylindrically outward to accelerate the equator \citep[][]{miesch2005,kapyla2011,hotta2014b,matilsky2020}.

It is well known that these convective columns propagate in a prograde direction faster than the local differential rotation speed \citep[][]{miesch2008,bessolaz2011}. This feature can be understood in terms of thermal Rossby waves \citep[e.g.,][]{busse2002}. They are prograde-propagating waves of $z$-vorticity (where $z$ denotes the rotational axis), originating from the conservation law of potential vorticity \citep[][]{miesch2005}. They owe their existence to both the topographic $\beta$-effect arising from the spherical curvature \citep[e.g.,][]{busse2002} and the compressional $\beta$-effect arising from the background density stratification \citep[][]{glatzmaier1981,evonuk2008,glatzmaier2009,verhoeven2014,ong2020}. In the Sun's convection zone, the compressional $\beta$-effect is dominant. Linear analyses of thermal Rossby modes were first carried out by \citet{glatzmaier1981} in the solar context, and later by \citet[][]{hindman2022,jain2023} using a simple one-dimensional model. A more sophisticated two-dimensional (2D) spherical shell model of the Sun's differentially rotating convection zone was recently presented by \citet[][]{bekki2022a}. Using fully nonlinear 3D convection simulations, \citet[][]{bekki2022b} further demonstrated that the thermal Rossby modes are the main transporters of enthalpy ($\approx 60\%$ of what is required) and angular momentum ($\approx 40\%$ of the net Reynolds stress) in the simulation. 
Despite their dominant presence and significant dynamical importance in the numerical models, these columnar convective patterns have never been successfully observed on the Sun. In the solar photospheric observations, the convective power declines monotonically above the supergranular scale \citep[e.g.,][]{hathaway2015}, implying the absence of large-scale convective columns. This puzzle is another manifestation of the convective conundrum.

A new promising explanation for how to generate the solar-like differential rotation was recently provided by \citet{hotta2022}, who showed that the small-scale Maxwell stress can transport the angular momentum radially upward against the downward transport by the Reynolds stress to help accelerate the equator. Although the role of the Maxwell stress in driving the differential rotation has been discussed by several authors \citep[e.g.,][]{fan2014,kapyla2017b,warnecke2025}, \citet{hotta2022}'s simulation was the first successful reproduction of solar-like differential rotation in a highly turbulent regime with the solar rotation rate $\Omega_{\odot}$ and luminosity $L_{\odot}$. It should be pointed out that, in their simulation, a low Ro is no longer needed to make the differential rotation solar-like. 
Complementary evidence was recently reported by \citet{soderlund2025}, who found a transition of rotational regime from anti-solar to solar-like at nearly fixed Ro by increasing the magnetic Prandtl number $\mathrm{Pm}$, thereby enhancing Lorentz forces relative to inertia.
This work, albeit in a Boussinesq framework, supports the view that magnetically dominated convection can favor solar-like differential rotation.

A key physical ingredient in \citet{hotta2022}'s simulation is an efficient small-scale dynamo (SSD), which produces super-equipartition magnetic fields at small scales \citep[see,][for a review on SSD]{rempel2023_review}. Although it has sometimes been doubted whether such an efficient SSD is possible in the Sun's convection zone where $\mathrm{Pm}$ is very small \citep[e.g.,][]{kapyla2018}, \citet{warnecke2023} recently showed that a SSD is possible in such a low-$\mathrm{Pm}$ regime. 
Since no current simulations have yet achieved the asymptotic regime of SSD with low $\mathrm{Pm}$, the actual efficiency of the SSD in the Sun is still unclear.
Nonetheless, given the enormous magnetic Reynolds number in the Sun convection zone, $\mathrm{Rm} \approx \mathcal{O}(10^{10})$ \citep[][]{ossendrijver2003}, it is very likely that the SSD operates more vigorously in the Sun than in any numerical models.
So far, most of the numerical investigations of SSD have been carried out in non-rotating systems \citep[e.g.,][]{rempel2014,hotta2015,kapyla2018,yan2021,bhatia2022}. Some numerical studies of rotating magneto-convection have reported the existence of SSD \citep[][]{favier2012,hotta2014b,hotta2016,warnecke2025} and the associated Maxwell stress which opposes the Reynolds stress \citep[][]{nelson2013,fan2014,augustson2015,kapyla2017b,warnecke2025}. However, few studies have focused on how SSD affects the properties of rotating columnar convection in the solar and stellar convection zones.

In this paper, we use a local Cartesian $f$-plane box model of rotating compressible convection at the equator to investigate the interaction between SSD and the columnar convective (thermal Rossby) modes. 
A similar local box model has been employed in previous studies of rotating magneto-convection \citep[][]{kapyla2009,favier2012,masada2014b,bushby2018}. In these studies, the box was located at the pole, with the rotational axis being parallel to the gravity.
Such a configuration provides helical turbulence, which can give rise to $\alpha^2$-type large-scale dynamo.
In this study, however, we exclude the $\alpha$-effect from our simulations by placing the $f$-plane box at the equator, which enables us to focus on the influence of the SSD on the columnar convective modes arising from the background density stratification.
Nevertheless, we note that this setup does not entirely rule out the possibility of large-scale dynamo action, because non-helical shear dynamos are also possible \citep[e.g.,][]{yousef2008,rogachevskii2003}.

The organization of the paper is as follows. In \S~\ref{sec:model}, our numerical model is explained. The results are presented in \S~\ref{sec:results}.
The conclusions are summarized in \S~\ref{sec:summary}.


\section{Numerical Model} \label{sec:model}

\subsection{Basic equations}

\begin{figure}
\begin{center}
\includegraphics[width=0.6\linewidth]{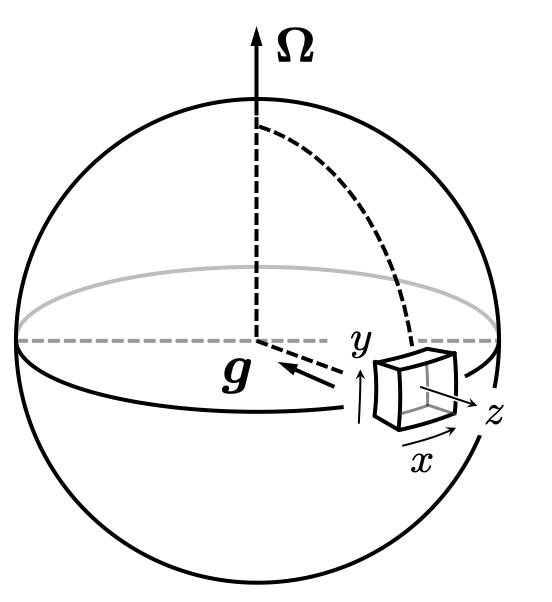}
\caption{
Sketch of the local Cartesian model of the stellar convection zone.
The local $f$-plane box is located at the equator, where the $x$, $y$, and $z$ axes point to the longitudinal, latitudinal, and radial directions, respectively.
}
\label{fig:cartesian}
\end{center}
\end{figure}

In this study, we numerically solve 3D magnetohydrodynamic (MHD) equations in a local Cartesian $f$-plane box. 
In our definition, the $x$-axis is directed from west to east (longitudinal), the $y$-axis from south to north (latitudinal), and the $z$-axis is directed radially upward.
The basic equations consist of the equation of continuity, the equation of motion, the induction equation, and the entropy equation
\begin{eqnarray}
  \frac{\partial \rho_1}{\partial t} &=& -\nabla\cdot [ (\rho_0+\rho_1)\bm{v} ], \label{eq:conti} \\
  \rho_0 \frac{\partial \bm{v}}{\partial t} &=& -\rho_0 \bm{v}\cdot\nabla\bm{v}-\nabla p_1
  +\rho_1 \bm{g}+2\rho_0 \bm{v}\times\bm{\Omega}_0  \nonumber \\
  && \ \ \ \ \ +\frac{1}{4\pi}(\nabla\times\bm{B})\times\bm{B} , \label{eq:motion} \\
  \frac{\partial \bm{B}}{\partial t} &=& \nabla\times(\bm{v}\times\bm{B}), \label{eq:induction} \\
  \rho_0 T_0 \frac{\partial s_1}{\partial t} &=& -\rho_0 T_0 \bm{v}\cdot\nabla s_1 +Q_{\mathrm{rad}}. \label{eq:ent} 
\end{eqnarray}
Here, $\rho_{0}$, $p_{0}$, and $T_{0}$ denote values of density, pressure, and temperature of a time-independent reference background stratification, respectively.
The reference background is assumed to be in an adiabatically stratified hydrostatic equilibrium with the gravitational acceleration $\bm{g}=-g\bm{e}_z$, where $g$ is assumed to be spatially constant.
The thermodynamic variables with subscript 1 ($\rho_{1}$, $p_{1}$, and $T_{1}$) represent perturbations that are small compared with the background values so that the equation of state can be linearized
\begin{eqnarray}
&& p_1 = p_0\left( \gamma\frac{\rho_1}{\rho_0}+\frac{s_1}{c_{\rm v}} \right). \label{eq:state}
\end{eqnarray}
For simplicity, we assume an ideal gas law with the specific heat ratio $\gamma = c_{\mathrm{p}}/c_{\mathrm{v}} = 5/3$ where $c_{\mathrm{p}}$ and $c_{\mathrm{v}}$ denote the specific heat at constant pressure and at constant volume, respectively.
In this paper, we locate the numerical box exactly at the equator so that the rotational axis is parallel to the $y$-axis, $\bm{\Omega}_0=\Omega_{0}\bm{e}_y$, as shown in Fig.~\ref{fig:cartesian}.

\subsection{Background stratification and radiative heating}

The time-independent background quantities $\rho_0$, $p_0$, and $T_0$ are given by an adiabatically stratified polytrope as \citep[e.g.,][]{fan1999}
\begin{eqnarray}
 &&  \rho_{0}(z)=\rho_{\mathrm{r}}\left[1-\frac{z}{(1+m)H_{\mathrm{r}}} \right]^{m}, \\
 && p_{0}(z)=p_{\mathrm{r}}\left[1-\frac{z}{(1+m)H_{\mathrm{r}}} \right]^{1+m}, \\
 && T_{0}(z)=T_{\mathrm{r}}\left[1-\frac{z}{(1+m)H_{\mathrm{r}}} \right], \\
 && H_{0}(z)=\frac{p_{0}}{\rho_{0} g},
\end{eqnarray}
where $m=1/(\gamma-1)$ is a polytropic index and $\rho_{\mathrm{r}}$, $p_{\mathrm{r}}$, $T_{\mathrm{r}}$, and $H_{\mathrm{r}}$ denote the values of $\rho_{0}$, $p_{0}$, $T_{0}$, and the pressure scale height $H_{0}$ at the bottom of our numerical domain $z = 0$.

In our model, the convective instability is driven by a prescribed radiative heating and cooling term through which the radiative energy flux is injected from the bottom boundary and extracted through the upper boundary;
\begin{eqnarray}
&& Q_{\mathrm{rad}}(z)=-\frac{\partial}{\partial z} [ F_{\mathrm{heat}}(z)+F_{\mathrm{cool}}(z) ].
\end{eqnarray}
We use the same functional forms for $F_{\mathrm{heat}}(z)$ and $F_{\mathrm{cool}}(z)$ as those presented in \citet{bekki2017b}.
The radiative heating part is assumed to be proportional to the background pressure as \citep{featherstone2016}
\begin{eqnarray}
  && -\frac{\partial F_{\mathrm{heat}}}{\partial z}= f_{\mathrm{heat}} [p_{0}(z)-p_{0}(z_{\mathrm{max}})], \label{eq:radheat}
\end{eqnarray}
where the normalization factor $f_{\mathrm{heat}}$ is determined by the input energy flux $F_{*}$ as
\begin{eqnarray}
  && f_{\mathrm{heat}}=F_{*}\left[\int_{0}^{z_{\mathrm{max}}} (p_{0}(z)-p_{0}(z_{\mathrm{max}})) dz \right]^{-1}.
\end{eqnarray}
The radiative cooling flux $F_{\mathrm{cool}}$ is assumed to be a Gaussian function localized near the surface.
\begin{eqnarray}
  &&F_{\mathrm{cool}}= F_{*} \exp{\left[ -\left(\frac{z-z_{\mathrm{max}}}{d_{\mathrm{cool}}} \right)^{2}\right]},
\end{eqnarray}
where the thickness of the surface cooling layer is set to be $d_{\mathrm{cool}}=0.3~H_{r}$, which is roughly the local pressure scale height at the top boundary.

\subsection{Dimensionless parameters}

The typical convective velocity $v_{*}$ is estimated as 
\begin{eqnarray}
   &&  v_{*}=\left( \frac{F_{*}}{\rho_r} \right)^{1/3}.
\end{eqnarray}
The typical convective turnover time scale $\tau_{*}$ and the typical magnetic field strength $B_{*}$ are then estimated respectively as 
\begin{eqnarray}
    && \tau_{*}=\frac{H_r}{v_{*}}, \ \ \ \ B_{*}=v_{*} \sqrt{4\pi\rho_r}. 
\end{eqnarray}
In our model, the radiative energy flux $F_{*}$ is set by specifying the modified Mach number $M_{*}$ (square root of the Euler number)
\begin{eqnarray}
    && M_{*}=\frac{v_{*}}{\sqrt{p_r/\rho_r}}.
\end{eqnarray}
Although $M_{*}$ is estimated to be $\approx \mathcal{O}(10^{-3}-10^{-4})$ in the Sun's convection zone \citep[][]{ossendrijver2003}, we use the value of $M_{*}=10^{-2}$ throughout this paper in order to avoid the severe CFL condition while keeping thermal convection sufficiently subsonic.
The rotational influence on convection is controlled by
\begin{eqnarray}
    && \mathrm{Co}_{*}=\frac{2\Omega_{0}H_{r}}{v_{*}},
\end{eqnarray}
which we vary as a free input parameter from $1$ to $20$ as shown in Table~\ref{table:1}.
We note that the parameter $\mathrm{Co}_{*}$ is defined \textit{a priori} on measurable quantities of the Sun (such as luminosity $L_{\odot}$, rotation rate $\Omega_{\odot}$) and is called a \textit{flux} Coriolis number \citep[e.g.,][]{aurnou2020,kapyla2024}.
For the readers' reference, the solar value of Co$_*$ is estimated to be $\mathrm{Co}_{\odot} \approx 3.8$ (with $F_{\odot}=1.2\times 10^{11}$~g~s$^{-3}$, $\Omega_{\odot}=2.8\times 10^{-6}$~s$^{-1}$, $H_r=57$~Mm, and $\rho_r = 0.2$~g~cm$^{-3}$).

\begin{table*}[]
 \begin{center} 
\caption{Summary of the numerical simulations reported in this paper.}
\small
\vspace{-0.3\baselineskip}
\begin{tabular}{cccccccccccccccccc} 
\toprule
\toprule
  \renewcommand{\arraystretch}{1.8}
Case & Co$_{*}$ & Co$_\ell$ & Re$_*$ & Rm$_*$ & $\Delta U_{x}$ & $\overline{v}_{\mathrm{rms}}$ & $\overline{\rho_0 s_{\rm rms}}$ & $\overline{\delta}$ & $\overline{k}_{x,\rm mean}$ & $\overline{k}_{x,\rm peak}$ & $E_{\mathrm{kin}}^{\prime}$ & $\widetilde{E}_{\mathrm{kin}}$ & $E_{\mathrm{mag}}$ & $\overline{\mathcal{R}}_{xz}$ & $\overline{\mathcal{M}}_{xz}$ \\
 \midrule
HD-Co1  & $1$  & 0.42 & 1087 & - & -0.75 & 2.07 &  2.13 &  1.98  & 22.6 & 1.17 & 2.17 & 0.35 & - & -0.03 & -  \\
HD-Co2  & $2$  & 0.93 & 1097 & - &-0.79 & 2.19 &  2.17 &  1.98  & 19.4 & 1.12 & 2.40 & 0.77 & - & 0.07 & -  \\
HD-Co5  & $5$  & 2.64 & 1084 & - &-2.99 & 2.45 &  2.18 &  2.09  & 15.3 & 1.92 & 2.88 & 1.25 & - & -0.03 & -  \\
HD-Co10 & $10$ & 4.05 & 1204 & - & 4.83 & 2.12 &  2.48 &  4.73  & 23.1 & 3.18 & 1.99 & 0.77 & - & 0.13 & -  \\
HD-Co20 & $20$ & 6.58 & 1450 & - &11.71 & 1.76 &  2.97 &  11.75 & 34.2 & 5.39 & 1.29 & 0.32 & - & 0.17 & -  \\
 \midrule
MHD-Co1  & $1$  & 0.68 & 1196 & 866 & -0.99 & 1.84 &  2.64 &  1.11 & 15.8 & 1.15 & 1.61 & 0.31 & 0.46 & -0.69 & 0.87   \\
MHD-Co2  & $2$  & 1.43 & 1180 & 917 & -2.15 & 1.85 &  2.66 &  1.07 & 15.0 & 1.26 & 1.60 & 0.49 & 0.51 & -1.96 & 2.12 \\
MHD-Co5  & $5$  & 3.93 & 1163 & 978 & -2.66 & 1.96 &  3.05 &  1.42 & 12.8 & 1.94 & 1.78 & 0.80 & 0.52 & -2.08 & 2.12 \\
MHD-Co10 & $10$ & 6.31 & 1144 & 1072 &  1.55 & 1.57 &  3.90 &  3.10 & 20.0 & 6.46 & 1.09 & 0.37 & 0.55 & 0.82 & -0.87 \\
MHD-Co20 & $20$ & 12.51 & 1050 & 1282 & 2.41 & 1.28 &  4.79 &  5.32 & 24.7 & 12.12 & 0.72 & 0.29 & 0.66 & 3.97 & -3.85 \\
\bottomrule
\end{tabular} \label{table:1}
\end{center}
\vspace{-1.0\baselineskip}
\tablefoot{
The flux Coriolis number Co$_*$ is the only explicit free parameter in this study, whereas the dynamical Coriolis number Co$_\ell$ is calculated as a simulation output.
The Reynolds and magnetic Reynolds numbers, Re$_*$ and Rm$_*$, depend on the grid resolution and are estimated in Appendix~\ref{appendix:numerical_diffusion}.
The amplitudes of the mean shear flow $\Delta U_{x}$ are shown in unit of $v_{*}$.
The volume-averaged values of rms convective velocity $\overline{v}_{\mathrm{rms}}$, rms entropy fluctuation $\overline{\rho_0 s_{\rm rms}}$, and superadiabaticity $\overline{\delta}$ are quoted in units of $v_{*}$, $\rho_r c_{\rm v} M_{*}^2$, and $M_{*}^{2}$, respectively.
The mean and peak longitudinal wavenumbers $\overline{k}_{x,\rm mean}$ and $\overline{k}_{x,\rm peak}$ are in unit of $2\pi L_x^{-1}$.
The volume-integrated kinetic energies of fluctuating velocity $E'_{\rm kin}=\int (\rho_0/2)\bm{v}'^2 dV$ and $y$-averaged velocity $\widetilde{E}_{\mathrm{kin}}$ and total magnetic energy $E_{\mathrm{mag}}$ are all quoted in unit of $10^2 \rho_r v_*^2 H_r^3$.
The mean values of the turbulent Reynolds and Maxwell stresses, $\overline{\mathcal{R}}_{xz}$ and $\overline{\mathcal{M}}_{xz}$, are reported in units of $\rho_r v_*^2$.
} 
\end{table*}

\subsection{Numerical methods}

We numerically solve Eqs.~(\ref{eq:conti})--(\ref{eq:ent}) using the 4th-order central-differencing method and 4-step Runge-Kutta scheme \citep[e.g.,][]{voegler2005}.
In order to keep the divergence-free condition of the magnetic field, the 9-wave method is used \citep{dedner2002}.
In this study, we do not use any explicit diffusivities but only use the slope-limited artificial diffusion proposed by \citet{rempel2014} to stabilize the computations (see Appendix~\ref{appendix:numerical_diffusion} for details). 
The kinetic and magnetic energy dissipated through this artificial diffusion is returned to the internal energy.
We note that this approach leads to effective thermal and magnetic Prandtl numbers in our simulations of order unity, both of which are much higher than those estimated in the real Sun.

The size of the numerical domain is $(L_{x},L_{y},L_{z})/H_{r}=(19.8, 6.6, 2.2)$, which is sufficiently large in the longitudinal ($x$) direction to capture the columnar convective pattern.
The density contrast between bottom and top of the numerical domain is about $24$ with the number of density scale heights $N_{\rho} \approx 3.2$.
We use a periodic boundary condition for horizontal directions and impenetrable, stress-free boundary condition at the vertical boundaries.
The magnetic field is assumed to be horizontal at the bottom ($z=0$) and vertical at the top boundary ($z=L_{z}$).
For all our simulations, we use the grid resolution of ($N_{x},N_{y},N_{z}$)=($1560$, $520$, $192$).
Owing to the nature of our diffusion scheme, this resolution determines the effective viscous, thermal, and magnetic diffusivities in the simulations.
Each simulation is initiated by giving a small random perturbation to the vertical velocity $v_z$ while setting the other hydrodynamic variables $\rho_1$, $v_x$, $v_y$, and $s_1$ to zero.
A weak seed field of $(B_x,B_y,B_z)/B_{*}=(0,10^{-5},0)$ is imposed to initiate the small-scale dynamo at $t=0$ in the MHD cases.

\begin{figure}
\begin{center}
\includegraphics[width=0.95\linewidth]{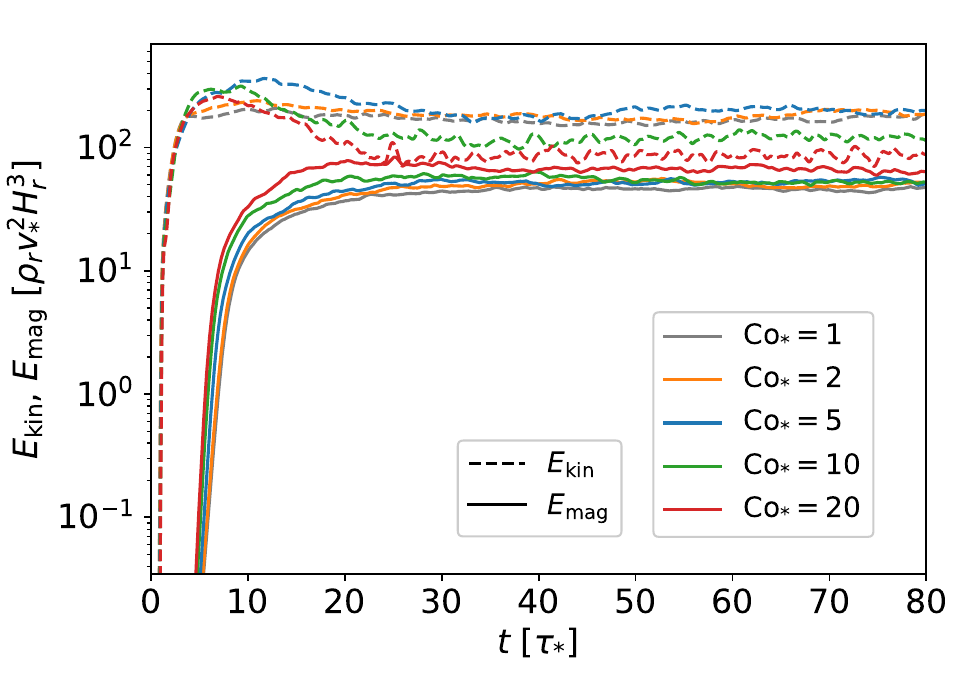}
\caption{
Temporal evolution of the volume-integrated kinetic and magnetic energies $E_{\mathrm{kin}}$ (dashed curves) and $E_{\mathrm{mag}}$ (solid curves) from the MHD runs.
Different colors represent the results with different Co$_{*}$.
}
\label{fig:EkinEmag_time}
\end{center}
\end{figure}

\begin{figure*}
\begin{center}
\includegraphics[width=0.99\linewidth]{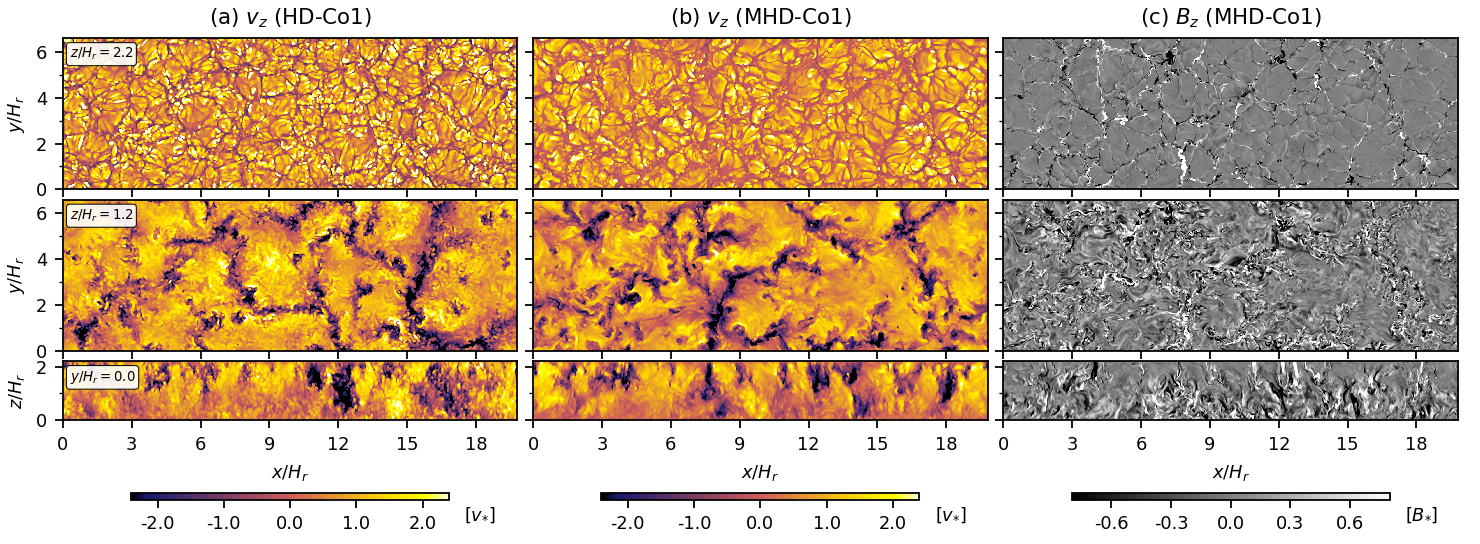}
\caption{
Temporal snapshots of (a) the vertical velocity $v_{z}$ from case HD-Co1, (b) $v_{z}$ from case MHD-Co1, and (c) the vertical magnetic field $B_{z}$ from MHD-Co1 in the statistically-stationary states at $t \approx 80~\tau_{*}$.
Top and middle panels show the horizontal cuts near the top surface, $z=2.2~H_{r}$, and in the middle convection zone $z=1.2~H_{r}$.
Bottom panels show the vertical cuts at $y=0$.
An animation of this figure is available \href{https://drive.google.com/file/d/1H_O0N2IodQlydZlGmBx8n7wW6egkAh91/view?usp=sharing}{online}.
}
\label{fig:snap_vz_Co1}
\end{center}
\end{figure*}
\begin{figure*}
\begin{center}
\includegraphics[width=0.99\linewidth]{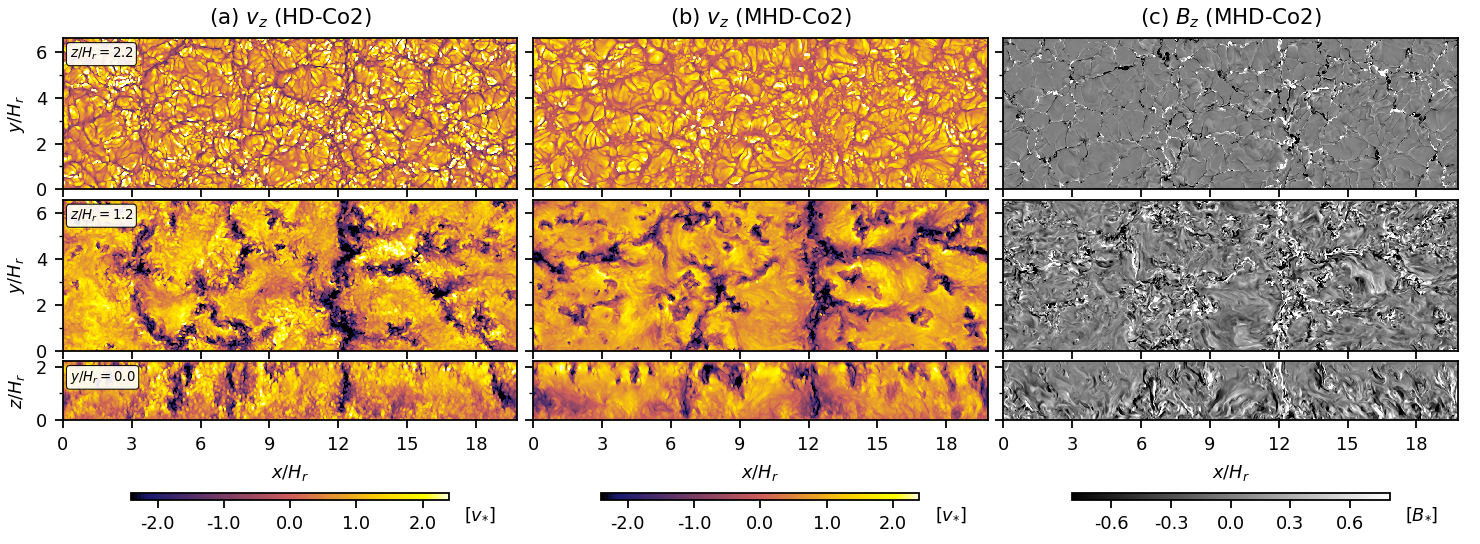}
\caption{
Same as Fig.~\ref{fig:snap_vz_Co1} but from the simulation cases HD-Co2 and MHD-Co2.
An animation of this figure is available \href{https://drive.google.com/file/d/10gmD3rqc8ZywXuIj2fhU0LfChBDvg4pd/view?usp=sharing}{online}.
}
\label{fig:snap_vz_Co2}
\end{center}
\end{figure*}
\begin{figure*}
\begin{center}
\includegraphics[width=0.99\linewidth]{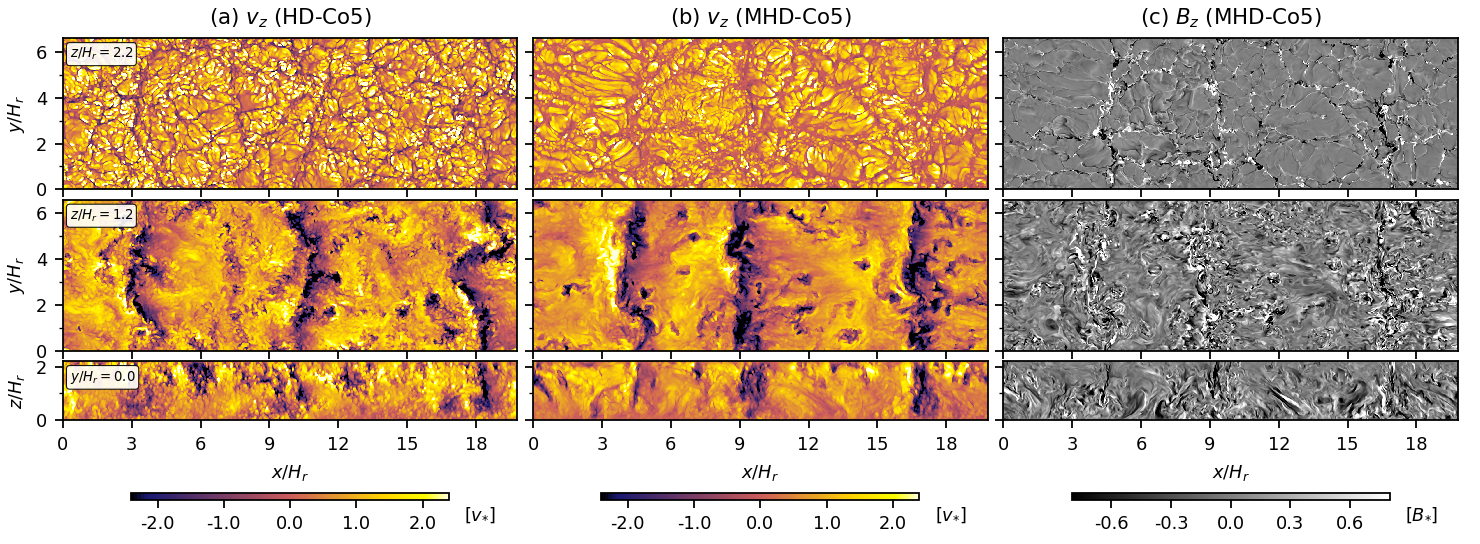}
\caption{
Same as Fig.~\ref{fig:snap_vz_Co1} but from the simulation cases HD-Co5 and MHD-Co5.
An animation of this figure is available \href{https://drive.google.com/file/d/1qMr4RnmgonlIrPpvW16R7COS9HJYSfZ9/view?usp=sharing}{online}.
}
\label{fig:snap_vz_Co5}
\end{center}
\end{figure*}
\begin{figure*}
\begin{center}
\includegraphics[width=0.99\linewidth]{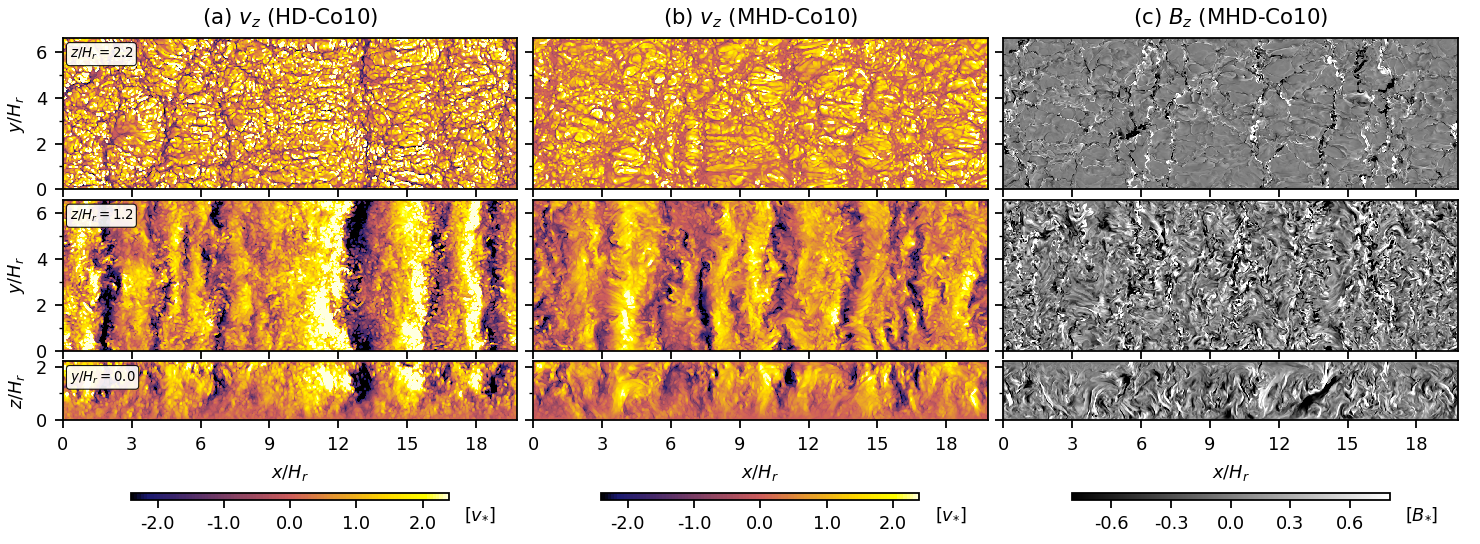}
\caption{
Same as Fig.~\ref{fig:snap_vz_Co1} but from the simulation cases HD-Co10 and MHD-Co10.
An animation of this figure is available \href{https://drive.google.com/file/d/14rVpxQZm_tUTa8T86d71FWhIOpZeyCBI/view?usp=sharing}{online}.
}
\label{fig:snap_vz_Co10}
\end{center}
\end{figure*}
\begin{figure*}
\begin{center}
\includegraphics[width=0.99\linewidth]{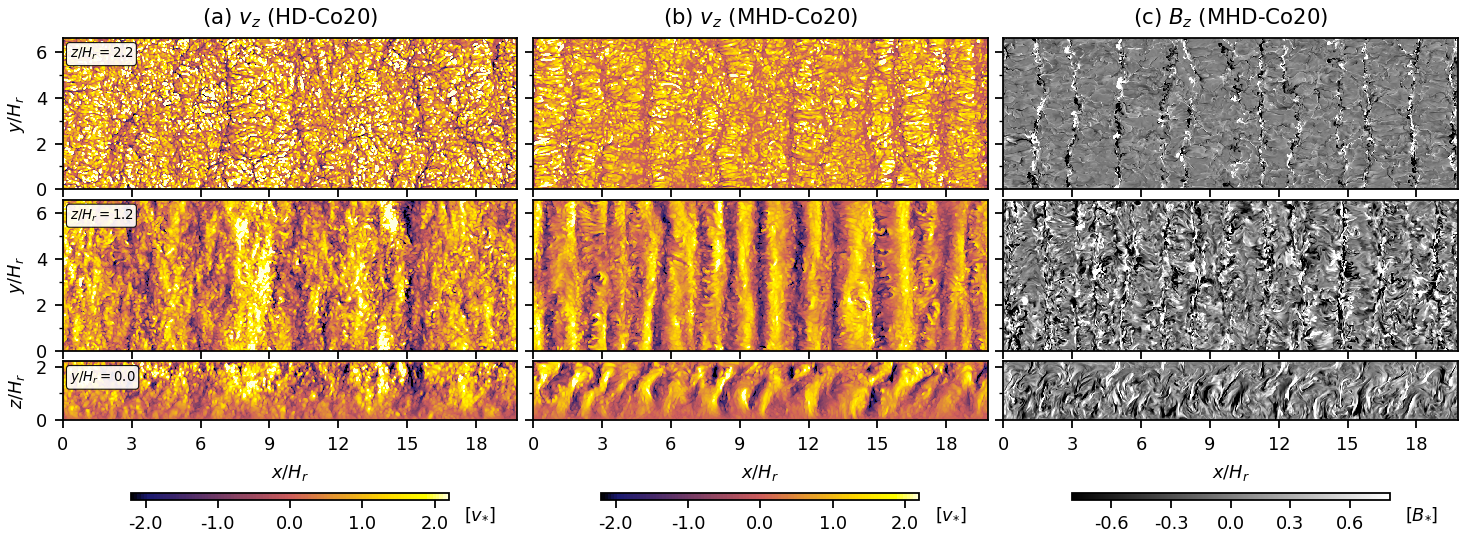}
\caption{
Same as Fig.~\ref{fig:snap_vz_Co1} but from the simulation cases HD-Co20 and MHD-Co20.
An animation of this figure is available \href{https://drive.google.com/file/d/1VFuNELaBGE0UwNuZVnJB3iOvZ-ocUntj/view?usp=sharing}{online}.
}
\label{fig:snap_vz_Co20}
\end{center}
\end{figure*}

\section{Results} \label{sec:results}

\subsection{Simulation overview} \label{sec:overview}

As summarized in Table~\ref{table:1}, we carry out sets of non-magnetic (denoted by "HD") and magnetic (denoted by "MHD") simulations with five different rotational influences (Co$_{*}=$1, 2, 5, 10, 20).
It is generally known that increasing Co$_*$ leads to a reduction in the supercriticality of convection \citep[e.g.,][]{chandrasekhar1961}. 
Exploring the effects of varying supercriticality requires huge numerical resources and is thus beyond the scope of this paper.
Since the $f$-plane box is located exactly at the equator, the mean flow is only established in $x$ (longitudinal) direction.
Furthermore, we note that our MHD simulations show no clear evidence of large-scale dynamo action.

Figure~\ref{fig:EkinEmag_time} shows the temporal evolution of the volume-integrated kinetic and magnetic energies $E_{\mathrm{kin}}$ and $E_{\mathrm{mag}}$ from the MHD cases.
They are defined by
\begin{eqnarray}
&& E_{\mathrm{kin}}=\int_V \frac{\rho_{0}}{2} (v_{x}^{2}+v_{y}^{2}+v_{z}^{2}) \ dV, \\
&& E_{\mathrm{mag}}=\int_V \frac{B_{x}^{2}+B_{y}^{2}+B_{z}^{2}}{8\pi} \ dV, \label{eq:Emag}
\end{eqnarray}
where $V = L_x L_y L_z$ is the total volume of the numerical domain.
All our simulations operate in a highly turbulent regime with effective Reynolds and magnetic Reynolds numbers, $\mathrm{Re}_{*},\ \mathrm{Rm}_{*} \approx 10^3$ (see Appendix~\ref{appendix:numerical_diffusion} for details), which are sufficiently large to ensure vigorous convection and efficient SSD action.
All the simulations are run for about $80$~$\tau_{*}$.
The SSDs are saturated and the statistically stationary states are reached by $t \approx 20$~$\tau_{*}$.
In what follows, we will focus on these statistically stationary states.

Figure~\ref{fig:snap_vz_Co1} shows the overall spatial patterns of vertical velocity $v_{z}$ and vertical magnetic field $B_{z}$ in the HD-Co1 and MHD-Co1 cases where the rotational influence is weak (Co$_*=1$).
In the MHD case (Fig.~\ref{fig:snap_vz_Co1}c), the results are very similar to those of non-rotating SSD simulations \citep[][]{cattaneo1999,voegler2005,rempel2014,hotta2015}.
Strong mixed-polarity magnetic fields are predominantly concentrated in the narrow downflow regions which are surrounded by slower and broader upflows.
Comparison between the HD and MHD cases (Figs.~\ref{fig:snap_vz_Co1}a and b) also reveals that the convective pattern in MHD becomes much smoother than that of HD simulation.
This can be attributed to the back-reaction from the SSD, i.e., the small-scale Lorentz force suppresses the small-scale turbulence \citep[][]{hotta2015}.

Figures~\ref{fig:snap_vz_Co2}, \ref{fig:snap_vz_Co5}, \ref{fig:snap_vz_Co10}, and \ref{fig:snap_vz_Co20} show the temporal snapshots of $v_{z}$ and $B_{z}$ in the same way as in Fig.~\ref{fig:snap_vz_Co1} but with increasing Co$_{*}$.
As Co$_*$ increases, the columnar pattern elongated in the $y$-direction becomes more and more apparent.
The size (longitudinal extent) of the convective columns becomes smaller with the increase in Co$_{*}$.
In other words, the total number of columns in the $x$-direction increases as convection becomes more and more rotationally constrained.
Furthermore, comparison between the HD and MHD cases again shows that the columnar pattern tends to be much more coherent and self-organized in the magnetized convection (e.g., Figs~\ref{fig:snap_vz_Co20}a and b).

\begin{figure*}    
\begin{center}
\includegraphics[width=0.85\linewidth]{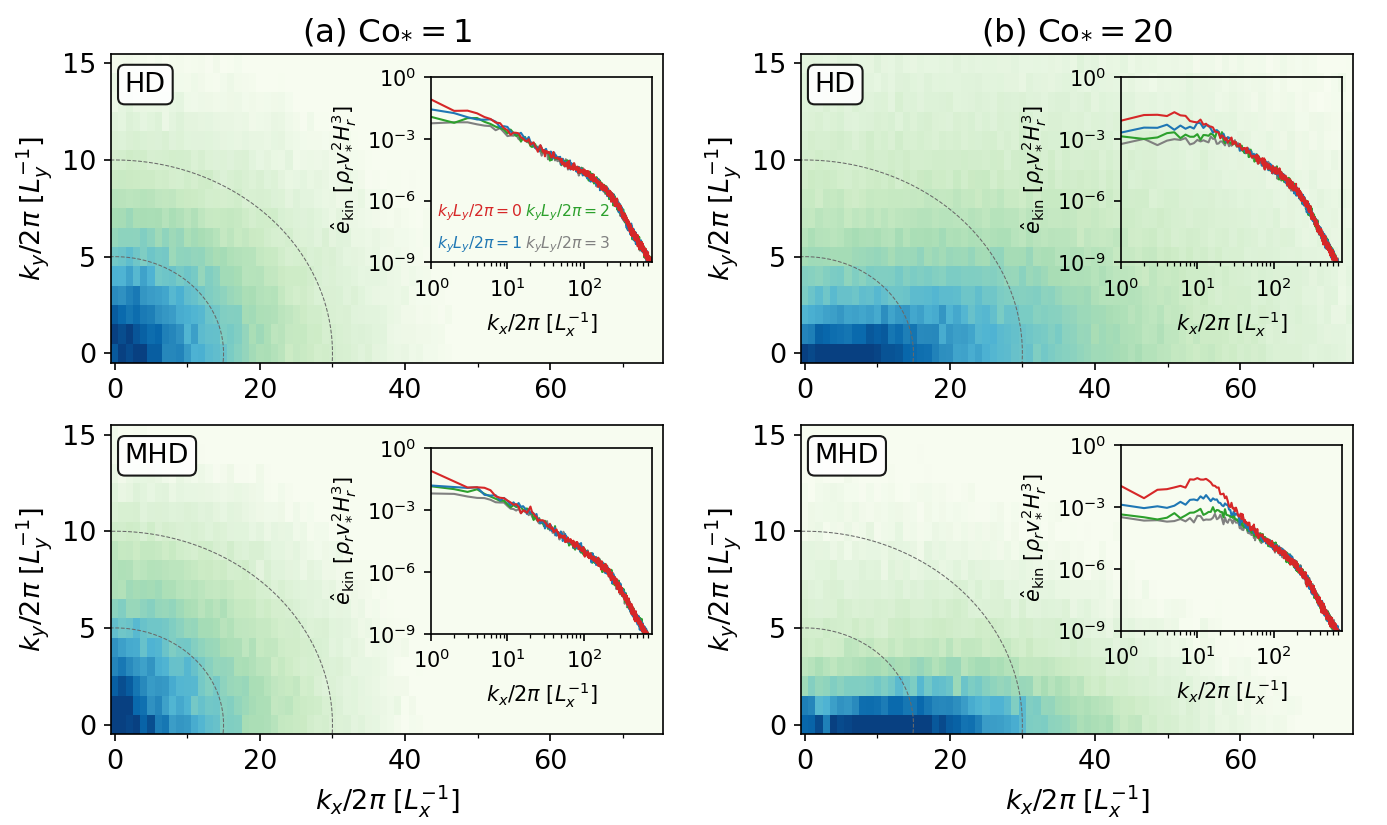}
\caption{
Two-dimensional kinetic energy spectra shown in a horizontal wavenumber space $(k_{x},k_{y})$ in the middle convection zone ($z=1.1~H_{r}$) for (a) weakly-rotating and (b) rapidly-rotating cases.
Top and bottom rows show the results of HD and MHD simulations, respectively.
The insets show cuts of the same spectra at fixed $k_{y}$, where the red, blue, green, and gray curves correspond to cuts through $k_{y}L_{y}/2\pi=0, \ 1, \ 2,$ and $3$.
}
\label{fig:Ekin_kxky}
\end{center}
\end{figure*}
\begin{figure*}    
\begin{center}
\includegraphics[width=0.95\linewidth]{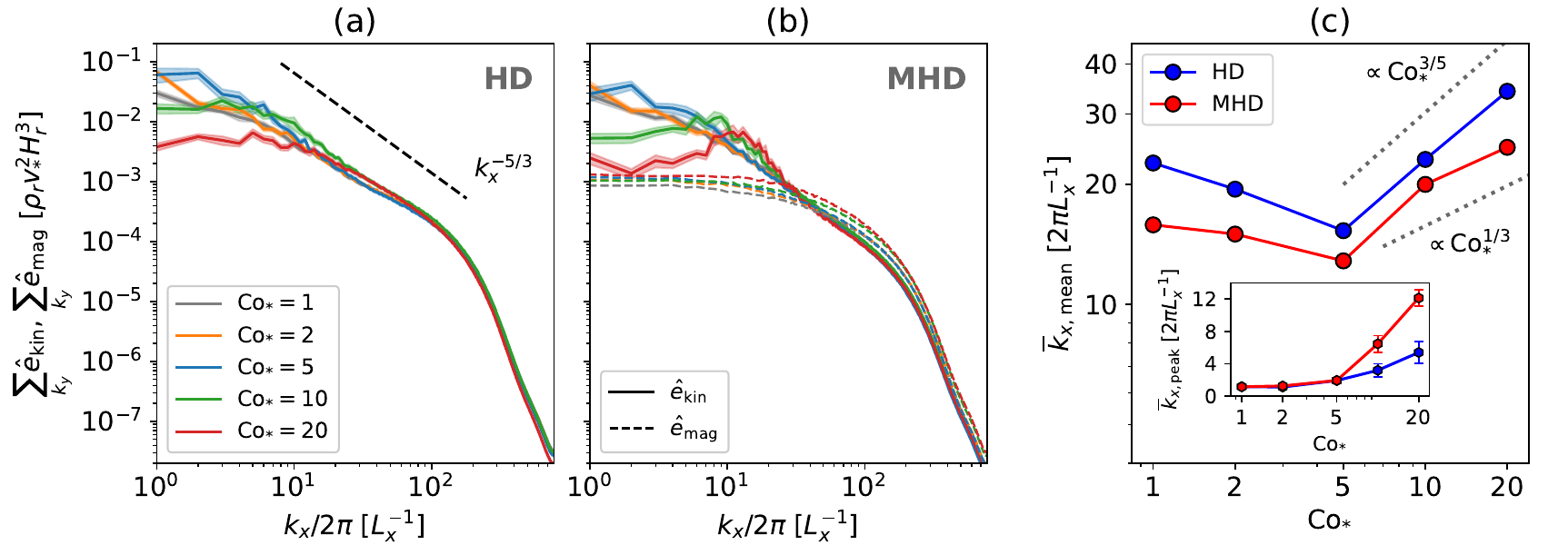}
\caption{
(a) Kinetic energy spectra $\sum_{k_y}\hat{e}_{\rm kin}$ as functions of longitudinal wavenumber $k_{x}$ in the middle convection zone ($z=1.1~H_{r}$) for HD cases.
Different colors correspond to different Co$_{*}$.
Black dotted line shows a $k_{x}^{-5/3}$ power law.
(b) The same kinetic energy spectra as panel (a) but for MHD cases.
Dotted curves show the magnetic energy spectra $\sum_{k_y}\hat{e}_{\rm mag}$ at the same depth.
(c) The mean wavenumber $\overline{k}_{x, \mathrm{mean}}$ defined by Eq.~(\ref{eq:kxmean}) as functions of Co$_{*}$. 
The blue and red colors correspond to the HD and MHD cases, respectively.
Gray dotted lines show $\propto \mathrm{Co}_{*}^{3/5}$ and $\propto \mathrm{Co}_{*}^{1/3}$ dependences theoretically expected from the CIA and MAC balances.
The inset shows the wavenumbers where the kinetic energy spectra peak, $\overline{k}_{x,\mathrm{peak}}$. 
}
\label{fig:Ekin_kmean}
\end{center}
\end{figure*}

\subsection{Notation of averages, Fourier spectra, and error estimates in this paper}

In the following sections, we discuss several statistical properties of rotating magneto-convection.
To this end, we define the following averaging procedures.
The horizontal ($x$ and $y$) average $\langle q \rangle$, the vertical ($z$) average $\overline{q}$, and the latitudinal ($y$) average $\widetilde{q}$ of a quantity $q$ are defined as
\begin{eqnarray}
&& \langle q \rangle (z) = \frac{1}{L_{x}L_{y}} \int_{0}^{L_{x}}\int_{0}^{L_{y}} q(x,y,z) dxdy, \\
&& \overline{q}(x,y) = \frac{1}{L_{z}}\int_{0}^{L_{z}} q(x,y,z) dz, \\
&& \widetilde{q}(x,z) = \frac{1}{L_{y}}\int_{0}^{L_{y}} q(x,y,z) dy.
\end{eqnarray}
Fluctuations with respect to the horizontally-averaged values are quoted by primes as
\begin{eqnarray}
    && q'(x,y,z)=q(x,y,z)-\langle q \rangle (z).
\end{eqnarray}
In addition to this, we explicitly denote the fluctuations with respect to the $y$-averaged quantities by $y$-primes as
\begin{eqnarray}
    && q^{\prime(y)}(x,y,z) = q(x,y,z) - \widetilde{q}(x,z).
\end{eqnarray}

We also define a Fourier transform in horizontal directions as
\begin{eqnarray}
    && q(x,y,z) = \sum_{k_x,k_y} \hat{q}(k_x,k_y,z) \ e^{\ii (k_x x + k_y y)} ,
\end{eqnarray}
where $k_x$ and $k_y$ are wavenumbers in the $x$ and $y$ directions.
We note that discretized wavenumbers $(k_x,k_y)=(2\pi n_x / L_x, 2\pi n_y/L_y)$ with integers $n_x,n_y \ (\ge 0)$ are used.

Hereafter, we will discuss the statistical properties averaged over time between $60 \leq t/\tau_{*} \leq 80$, with the standard errors given by $\sigma/\sqrt{N_{t}-1}$ where $\sigma$ denotes the standard deviation and $N_{t}$ is the number of temporal snapshots used.
To ensure that the snapshots are statistically independent, we use the data at intervals of $1.0~\tau_{*}$, i.e., $N_{t}=20$.

\subsection{Energy spectra} \label{sec:spectra}

\begin{figure*}
\begin{center}
\includegraphics[width=0.8\linewidth]{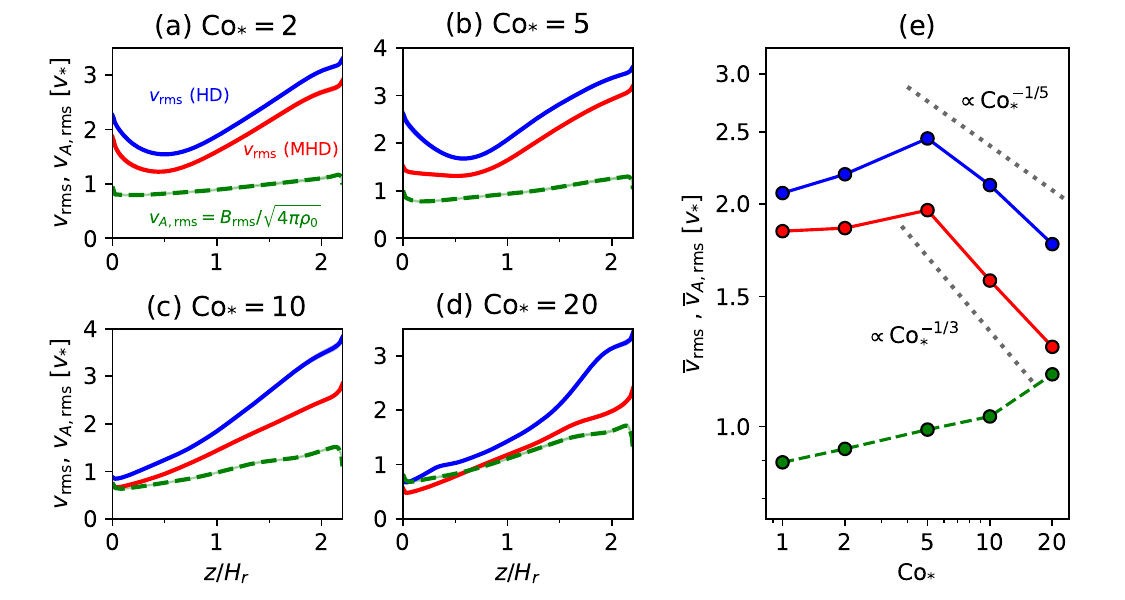}
\caption{
Profiles of the root-mean-square (rms) convective velocity $v_{\rm rms}$ and the Alfvén velocity of the rms magnetic field $v_{\rm A, rms}=B_{\rm rms}/\sqrt{4\pi\rho_{0}}$.
(a--d) Vertical profiles of $v_{\rm rms}$ for different $\mathrm{Co}_{*}$ where blue and red represent the results from HD and MHD cases.
Green dashed curves show the profiles of $v_{\rm A, rms}$ from MHD cases.
(e) The values of volume-averaged rms velocity and magnetic field, $\overline{v}_{\rm rms}$ and $\overline{v}_{\rm A, rms}$, as a function of $\mathrm{Co}_{*}$.
Gray dotted lines show $\propto \mathrm{Co}_{*}^{-1/5}$ and $\propto \mathrm{Co}_{*}^{-1/3}$ dependences theoretically expected from the CIA and MAC balances.
}
\label{fig:vrmsBrms}
\end{center}
\end{figure*}

Figure~\ref{fig:Ekin_kxky} exemplifies the 2D kinetic energy spectra in the middle convection zone ($z=1.1~H_{r}$) for weak rotation (Co$_{*}=1$) and strong rotation (Co$_{*}=20$) cases.
Here, the kinetic energy spectrum $\hat{e}_{\rm kin}$ is defined as
\begin{eqnarray}
    && \hat{e}_{\rm kin}(\bm{k},z) = \frac{\rho_0}{2} \hat{\bm{v}}(\bm{k},z) \cdot \hat{\bm{v}}^{*}(\bm{k},z).
\end{eqnarray}
By definition, this satisfies
\begin{eqnarray}
    && \Bigl \langle \frac{\rho_0}{2} \bm{v}^{2} \Bigr \rangle (z) = \sum_{\bm{k}} \hat{e}_{\rm kin}(\bm{k},z).
\end{eqnarray}
Although the distribution of convective power in wavenumber space $(k_x,k_y)$ is almost isotropic in Co$_{*}=1$ cases, the spectra become strongly anisotropic in Co$_{*}=20$ cases wherein most power is concentrated in the Fourier domain where the latitudinal wavenumber $k_y$ is small.
In fact, as indicated by the inset panels in Fig.~\ref{fig:Ekin_kxky}b, the $k_y=0$ component of the convective motions is shown to be dominant over the $k_y \ne 0$ components.
This corresponds to a formation of coherent columnar patterns aligned in the $y$-direction (Figs.~\ref{fig:snap_vz_Co10} and \ref{fig:snap_vz_Co20}).

Figures~\ref{fig:Ekin_kmean}a and b show the kinetic energy spectra summed over $k_y$ for HD and MHD cases for different values of Co$_*$.
When Co$_*$ is small, the kinetic energy spectra roughly follow the Kolmogorov power law $\propto k_{x}^{-5/3}$ in the inertial spectral range.
However, when Co$_*$ is increased, $\hat{e}_{\rm kin}$ begins to show spectral peaks at higher longitudinal wavenumbers $k_{x}$.
In MHD simulations, we also show the magnetic energy spectra $\hat{e}_{\rm mag}$ defined by
\begin{eqnarray}
    && \hat{e}_{\rm mag}(\bm{k},z) = \frac{1}{8\pi} \hat{\bm{B}}(\bm{k},z) \cdot \hat{\bm{B}}^{*}(\bm{k},z).
\end{eqnarray}
We find that the SSD generates super-equipartition (i.e., $\hat{e}_{\rm kin} \lesssim \hat{e}_{\rm mag}$) magnetic fields at small scales.
Consequently, the kinetic energy $\hat{e}_{\rm kin}$ is suppressed due to the Lorentz force feedback at these scales.

It is important to note that, in our simulations, there are two distinct characteristic length scales:
One is the mean longitudinal wavenumber $k_{x,\rm mean}$ weighted by the kinetic energy spectrum \citep[e.g.,][]{christensen2006,kapyla2024},
\begin{eqnarray}
    && k_{x,\mathrm{mean}}(z) = \frac{\sum_{\bm{k}} k_{x} \hat{e}_{\rm kin}(\bm{k},z) }{\sum_{\bm{k}} \hat{e}_{\rm kin}(\bm{k},z) }, \label{eq:kxmean}
\end{eqnarray}
and the other is the peak longitudinal wavenumber $k_{x,\rm peak}$ at which the kinetic energy spectrum reaches its local maximum. 
In previous studies of rotating $f$-plane simulations where gravity is aligned with the rotation axis, these two wavenumbers become asymptotically equivalent in the rapidly rotating limit \citep[e.g.,][]{kapyla2024}. 
However, in our model setup, where both small-scale convective motions and large-scale columnar convective modes coexist, they do not coincide.
Figure~\ref{fig:Ekin_kmean}c shows the height-averaged mean wavenumber $\overline{k}_{x,\rm mean}$ and the peak wavenumber $\overline{k}_{x,\rm peak}$ as functions of Co$_*$. 
In general, the inclusion of the SSD leads to smaller $\overline{k}_{x,\rm mean}$ in MHD (i.e., the typical convective scale $\overline{\ell}_{x}=2\pi/\overline{k}_{x,\rm mean}$ becomes larger) than in HD because the SSD Lorentz force suppresses small-scale turbulence.
In contrast, the peak wavenumber $\overline{k}_{x,\rm peak}$ is shown to be systematically larger in MHD at Co$_* \ge 5$.
We find that the peak wavenumber $\overline{k}_{x,\rm peak}$ largely represents the dominant scale of the $k_y=0$ columnar convective (thermal Rossby) modes.
The effects of the SSD on $\overline{k}_{x,\rm peak}$ will be discussed later in \S~\ref{sec:invcascade}.

\subsection{Convective heat transport} \label{sec:vrms_delta}

\begin{figure*}
\begin{center}
\includegraphics[width=0.975\linewidth]{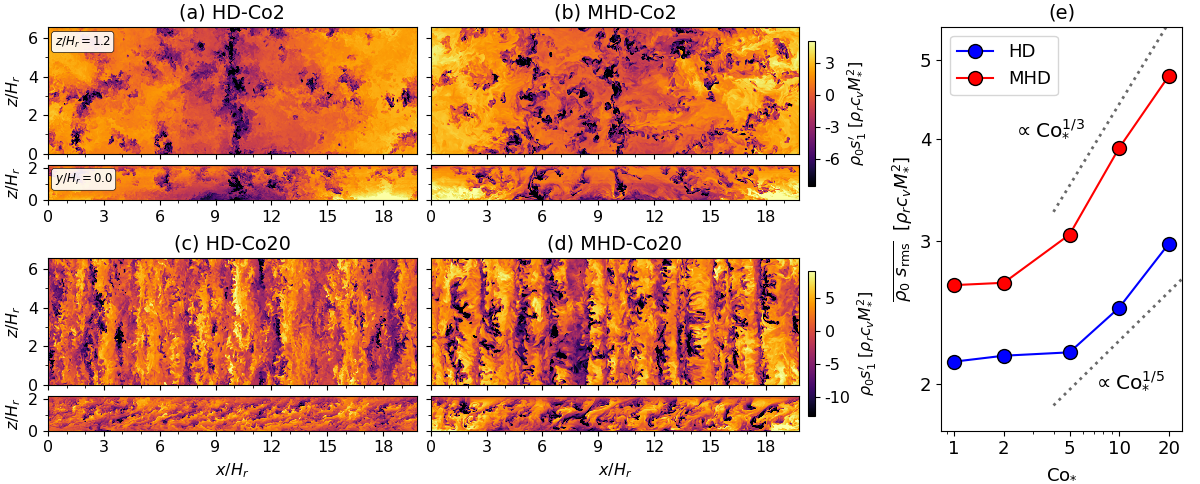}
\caption{
Entropy perturbations $s'_1$ in our simulations.
(a,b) Temporal snapshots of $s'_1$ weighted by the background density $\rho_0$ at fixed height ($z=1.2~H_r$) and latitude ($y=0$) from weakly rotating simulations HD-Co2 and MHD-Co2.
(c,d) Same as panels a and b but from rapidly rotating cases HD-Co20 and MHD-Co20.
(e) The volume-averaged rms entropy perturbation, $\overline{\rho_0 s_{\rm rms}}$, as a function of Co$_{*}$.
Blue and red denote the results from HD and MHD simulations.
Gray dotted lines show $\propto \mathrm{Co}_{*}^{1/5}$ and $\propto \mathrm{Co}_{*}^{1/3}$ dependences theoretically expected from the CIA and MAC balances.
}
\label{fig:s1rms}
\end{center}
\end{figure*}

Figures~\ref{fig:vrmsBrms}a--d show vertical profiles of the root-mean-square (rms) convective velocity
\begin{eqnarray}
    && v_{\rm rms}(z) = \sqrt{\langle (\bm{v}(x,y,z)-U_x(z) \bm{e}_x)^2 \rangle},
\end{eqnarray}
where $U_x = \langle v_x \rangle$ denotes the mean flow in the $x$-direction.
Generation mechanisms of the mean flows will be discussed in a later section (\S~\ref{sec:meanflows}).
From MHD simulations, we also show the Alfvén velocity associated with the rms magnetic field
\begin{eqnarray}
    && v_{\rm A, rms}(z) = \sqrt{\frac{{\langle \bm{B}(x,y,z)^2 \rangle}}{{4\pi \rho_0 (z)}}}. 
\end{eqnarray}
It is evident that, due to the SSD Lorentz force feedback, $v_{\rm rms}$ in the MHD cases is suppressed by approximately $15$--$25$\% compared to the HD cases.
In most of our simulations, the rms field strength $B_{\rm rms}$ is sub-equipartition, i.e., $v_{\rm A, rms} < v_{\rm rms}$.
Although this is in accordance with the simulations by \citet{favier2012} and \citet{warnecke2025}, we note that the higher-resolution simulation by \citet{hotta2022} produces super-equipartition SSD fields throughout the whole convection zone.

It is generally known that strong rotation tends to suppress convective velocities \citep[e.g.,][]{aurnou2020, vasil2021, kapyla2024}. 
As shown in Fig.~\ref{fig:vrmsBrms}e, our simulations also demonstrate that the volume-averaged rms convective velocity $\overline{v}_{\rm rms}$ decreases with increasing Co$_*$ in the rapidly rotating regime (Co$_* \ge 5$) in both HD and MHD cases. 
Moreover, our study reveals that this suppression is further enhanced by the presence of SSD. 
In HD, $\overline{v}_{\rm rms}$ decreases gradually with Co$_*$, roughly following a power law of $\mathrm{Co}_*^{-1/5}$, while in MHD, the decline is steeper, following $\mathrm{Co}_*^{-1/3}$, as indicated by the gray dotted lines in Fig.~\ref{fig:vrmsBrms}e. 
The underlying mechanism by which the SSD modifies the Co$_*$ dependence will be explained in \S~\ref{sec:force}.

Next, we discuss thermal energy transport properties in our simulations.
Figures~\ref{fig:s1rms}a--d present snapshots of the entropy perturbations $\rho_0 s'_{1}$ for Co$_* =2$ and $20$.
Figure~\ref{fig:s1rms}e further shows the vertically-averaged rms entropy perturbations $\overline{\rho_0 s_{\rm rms}}$ as a function of Co$_*$.
Here, $s_{\rm rms}$ is defined by 
\begin{eqnarray}
  &&  s_{\rm rms}(z) = \sqrt{\langle s_1^{\prime}(x,y,z)^2 \rangle}.
\end{eqnarray}
We note that the entropy perturbations shown are weighted by the background density $\rho_0$, which allows us to focus on the entropy variations within the bulk convection zone, rather than those generated by strong near-surface radiative cooling.
Two main effects are observed. First, we find an enhancement of the entropy perturbations $s'_1$ due to the SSD action. 
This behavior was reported in earlier non-rotating studies \citep[e.g.,][]{hotta2015}, where it was shown that the SSD magnetic fields suppress horizontal turbulent mixing of entropy between warm upflows and cold downflows, thereby enhancing entropy contrast.
At small Co$_*$, the MHD simulations show an increase in $s_{\rm rms}$ by approximately $25$\% compared to the HD cases.
The second effect is an enhancement of $s'_1$ by strong rotation at Co$_* \ge 5$. It is well known that the Coriolis force makes the convective heat transport inefficient as it deflects vertical motions and reduces the entropy mixing in a vertical direction. 
To sustain a fixed enthalpy flux in the domain, the entropy perturbations $s'_1$ are enhanced.
As shown in Fig.~\ref{fig:s1rms}e, the volume-averaged rms entropy perturbation $\overline{\rho_0 s_{\rm rms}}$ increases with Co$_*$ in both HD and MHD cases. 
However, the scaling differs: the MHD cases follow an approximate scaling of $\mathrm{Co}_*^{1/3}$, while the HD cases follow $\mathrm{Co}_*^{1/5}$, indicating a stronger Co$_*$ dependence with the presence of SSD.
Obviously, these scalings are the inverse of those found in the rms convective velocity $v_{\rm rms}$ (Fig.~\ref{fig:vrmsBrms}e), implying that the entropy perturbations are enhanced to compensate for the reduced convective velocities and to transport the fixed amount of enthalpy upward.

\begin{figure*}
\begin{center}
\includegraphics[width=0.8\linewidth]{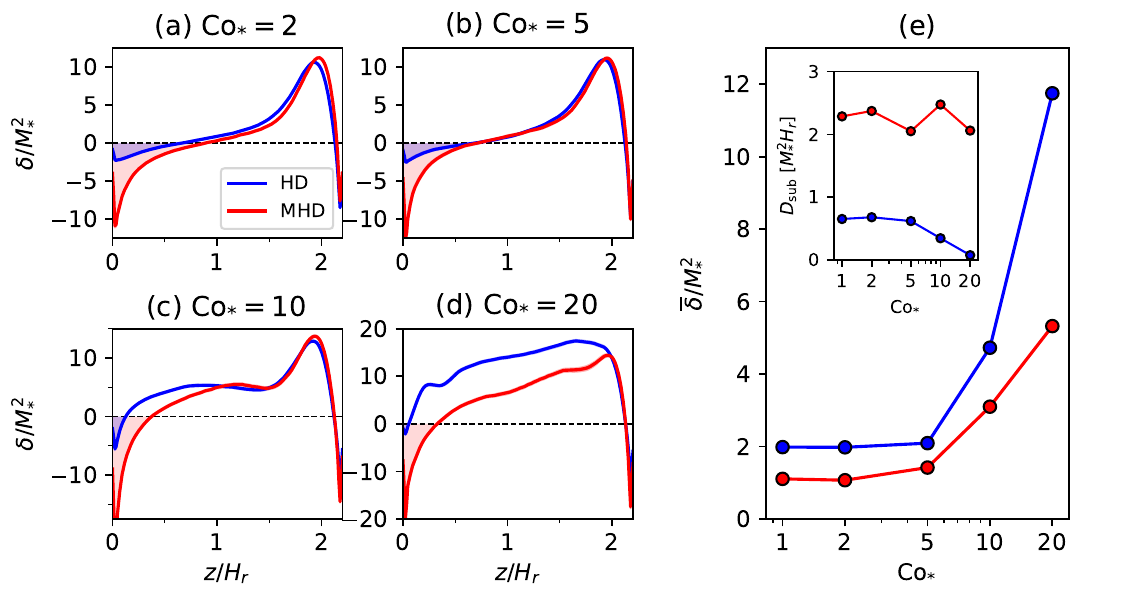}
\caption{
Profiles of the superadiabaticity $\delta$.
(a--d) Vertical profiles of $\delta$ for different $\mathrm{Co}_{*}$.
Blue and red curves represent the results from HD and MHD simulations.
(e) The volume-averaged values of superadiabaticity $\overline{\delta}$ as a function of $\mathrm{Co}_{*}$.
The inset shows the magnitudes of the subadiabatic layers near the bottom of the convection zone $D_{\rm sub}=\int_0^{d_{\rm sub}} |\delta|dz$ (with $d_{\rm sub}$ being the height below which the mean stratification is subadiabatic) denoted by the shaded areas in panels a--d.
}
\label{fig:delta}
\end{center}
\end{figure*}

The efficiency of convective energy transport is known to be associated with the mean entropy stratification in the bulk convection zone. Figures~\ref{fig:delta}a--d show vertical profiles of the superadiabaticity $\delta$ defined as
\begin{eqnarray}
    && \delta = -\frac{H_0}{c_{\rm p}}\frac{\partial \langle s_1 \rangle}{\partial z}. \label{eq:delta}
\end{eqnarray}
When the rotational influence is weak, a weakly subadiabatic layer is formed in the lower half of the convection zone. This is because of the non-local heat transport by strong downflow plumes \citep[e.g.,][]{brandenburg2016} and has been repeatedly reported in previous simulations of non-rotating convection \citep[e.g.,][]{kapyla2017,bekki2017b,karak2018,kapyla2025}. It is also seen that the magnitude and width of this subadiabatic layer are enhanced and extended with the inclusion of SSD. When Co$_{*}$ is increased, the mean stratification becomes more superadiabatic in the bulk convection zone because the rotation suppresses vertical entropy mixing.
These trends are clearly seen in Fig.~\ref{fig:delta}e. Interestingly, we find that the SSD makes the subadiabatic layer near the base of the convection zone persistent even under strong rotational influence (see the inset of Fig.~\ref{fig:delta}e).
This result is in striking contrast to the HD case where the subadiabatic layer vanishes in the rapidly rotating regime.
A disappearance of the subadiabatic layer in the rapidly rotating HD regime has also been reported by \citet{kapyla2024}, who employed a radiative diffusivity based on Kramers opacity (as a function of temperature and density) instead of the spatially fixed radiative heating used in this study.

Our parameter study suggests that the weakly subadiabatic layer likely exists in the lower $40\%$ of the Sun's convection zone near the equator (given that Co$_{\odot} \approx 3.8$). 
A similar result is obtained by \citet[][]{hotta2022}.
We also note that the existence of this weakly subadiabatic layer in the Sun's convection zone has been inferred by recent observations and linear modeling of the non-toroidal class of the solar inertial modes \citep{hanson2022,bekki2024a}.


\subsection{Force balance} \label{sec:force}

\begin{figure*}
\begin{center}
\includegraphics[width=0.95\linewidth]{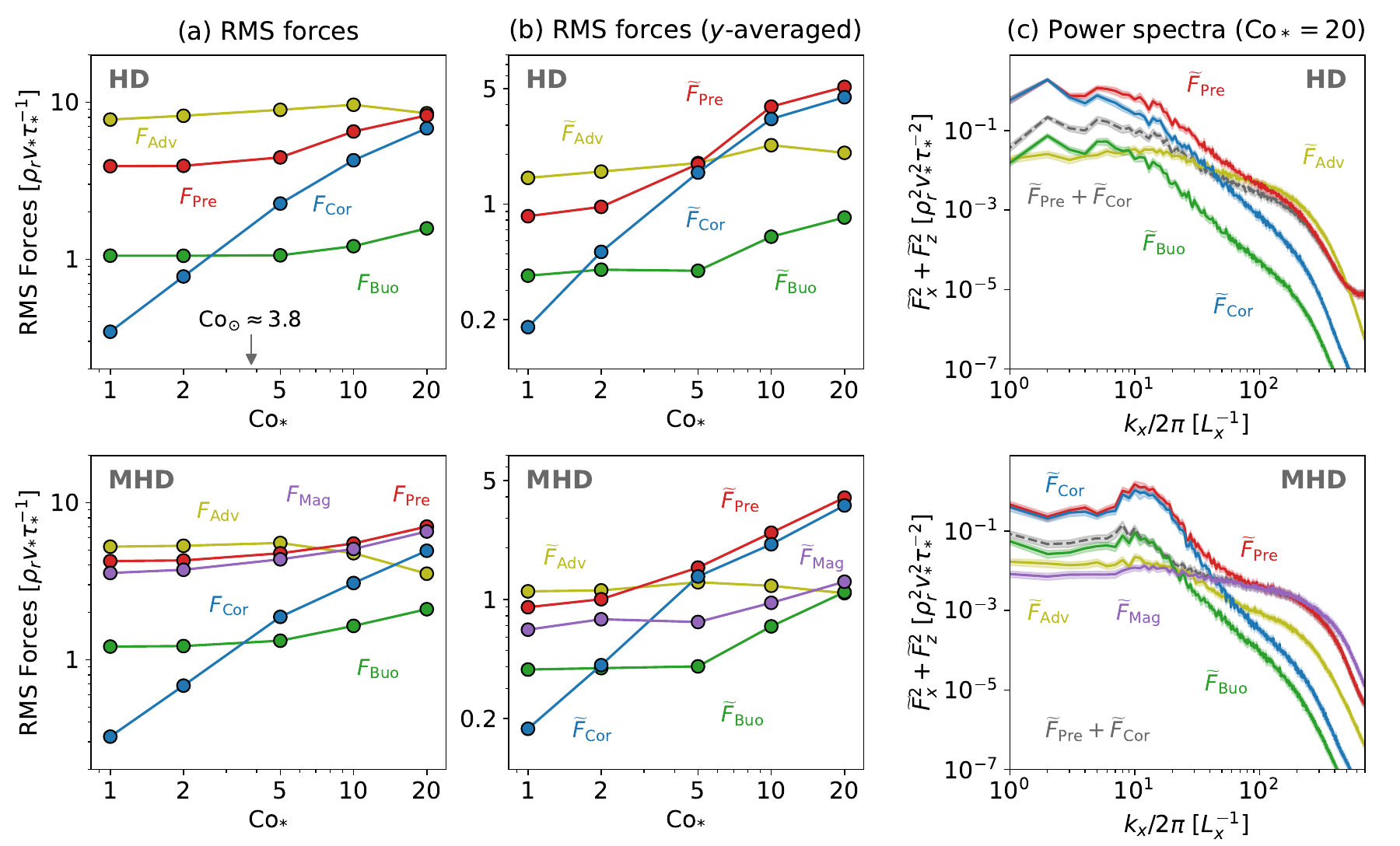}
\caption{
Force balances in our simulations.
(a) Root-mean-square (rms) amplitudes of the forces $F_{\rm rms}$.
Here, $F_{\rm rms}$ is calculated such that $F_{\rm rms}^2 = V^{-1}  \int_{V} (F_x^{\prime 2} + F_y^{\prime 2} + F_z^{\prime 2}) dV$.
Yellow, red, green, blue, and purple colors represent the nonlinear advection, pressure gradient, buoyancy, Coriolis, and Lorentz forces.
Top and bottom panels show the HD and MHD simulations, respectively.
(b) RMS amplitudes of the $y$-averaged forces $\widetilde{F}_{\rm rms}$.
They are defined such that $\widetilde{F}_{\rm rms}^2 = V^{-1}  \int_{V} (\widetilde{F}_x^{\prime 2} + \widetilde{F}_z^{\prime 2}) dV$.
(c) Power spectra of the $y$-averaged forces in the middle height $z=1.1~H_{r}$ from HD-Co20 and from MHD-Co20 simulations.
Gray dashed curves denote the ageostrophic Coriolis forces $\widetilde{F}_{\rm Pre}+\widetilde{F}_{\rm Cor}$.
}
\label{fig:Forces_mac}
\end{center}
\end{figure*}

To understand why convective velocity $v_{\rm rms}$ and entropy fluctuation $s_{\rm rms}$ show different Co$_*$ dependences with and without the SSD, we investigate the dominant force balance in our simulations.
We follow the method presented in \citet[][]{yadav2016,guzman2021} and compute the rms values of the following forces
\begin{eqnarray}
    && F_{\rm Adv}=-\rho_{0}\bm{v}\cdot\nabla\bm{v},  \label{eq:force_adv}\\
    && F_{\rm Pre}=-\rho_{0}\nabla\left(\frac{p_{1}}{\rho_{0}} \right), \label{eq:force_pre}\\
    && F_{\rm Buo}= \rho_{0}\frac{s_{1}}{c_{p}} g, \label{eq:force_buo}\\
    && F_{\rm Cor}= 2\rho_{0}\bm{v} \times \bm{\Omega}_{0}, \label{eq:force_cor} \\
    && F_{\rm Mag}= \frac{1}{4\pi} (\nabla\times\bm{B})\times \bm{B}. \label{eq:force_mag}
\end{eqnarray}
Here, we note that temporally-averaged component of the pressure and buoyancy forces (which represents the background hydrostatic balance and thus is irrelevant to our analysis) are subtracted from $F_{\rm Pre}$ and $F_{\rm Buo}$.
Figure~\ref{fig:Forces_mac}a shows the rms forces in HD (top) and MHD (bottom) cases as functions of Co$_*$.
In HD simulations, the advection (or inertia) term $F_{\rm Adv}$ always dominates the force balance.
As Co$_*$ is increased, a balance is achieved among advection $F_{\rm Adv}$, pressure $F_{\rm Pre}$, and Coriolis forces $F_{\rm Cor}$.
On the other hand, in MHD simulations, the Lorentz force $F_{\rm Mag}$ comes into play in the leading-order force balance, where $F_{\rm Mag}$ has amplitudes very close to those of the pressure gradient force $F_{\rm Pre}$.
This likely reflects the fact that the strong SSD fields are generated by compression of downflows where the gas pressure is locally balanced by the magnetic pressure \citep{hotta2022}.
At sufficiently high Co$_*$, $F_{\rm Mag}$ is even stronger than $F_{\rm Adv}$, leading to a balance between pressure $F_{\rm Pre}$, Coriolis $F_{\rm Cor}$, and the Lorentz forces $F_{\rm Mag}$.
This balance is called \textit{magnetostrophic} \citep[e.g.,][]{roberts1978}.
We can attribute the Co$_*$ dependences of $v_{\rm rms}$ and $s_{\rm rms}$ discussed in \S~\ref{sec:vrms_delta} to this transition of the dominant force balance regime.

It is worth noting that the relative importance of the Coriolis force $F_{\rm Cor}$ (which does not have a $y$-component) tends to be underestimated in Fig.~\ref{fig:Forces_mac}a.
To properly assess the significance of the Coriolis force in the system, we also compute the same rms forces but for the $y$-averaged ones, as shown in Fig.~\ref{fig:Forces_mac}b.
We can see that there is a clear distinction between the weakly rotating regime Co$_* < 5$ where the Coriolis force $\widetilde{F}_{\rm Cor}$ is not in the leading-order force balance and the strongly rotating regime Co$_* \gtrsim 5$ where the dominant force balance is achieved between the Coriolis force $\widetilde{F}_{\rm Cor}$ and the pressure gradient force $\widetilde{F}_{\rm Pre}$.
When this geostrophic balance is established at a leading order, it is necessary to look at the residual balance \citep[e.g.,][]{yadav2016,aubert2017}.
Figure~\ref{fig:Forces_mac}c displays the power spectra of the $y$-averaged forces in the middle convection zone ($z=1.1~H_{r}$) from rapidly-rotating HD-Co20 and MHD-Co20 cases. 
Here, the imbalance between Coriolis and pressure gradient forces, $\widetilde{F}_{\rm Cor}+\widetilde{F}_{\rm Pre}$, are denoted by gray dashed curves.
In the HD case, this so-called \textit{ageostrophic} Coriolis force is balanced by buoyancy $\widetilde{F}_{\rm Buo}$ and advection $\widetilde{F}_{\rm Adv}$ at large scales and predominantly by advection at small scales, manifesting the Coriolis-Inertia-Archimedean (CIA) balance \citep[see, e.g.,][]{aurnou2020,kapyla2024}.
In the MHD case, on the other hand, the ageostrophic Coriolis force is primarily balanced by buoyancy $\widetilde{F}_{\rm Buo}$ at large scales but is balanced by Lorentz force $\widetilde{F}_{\rm Mag}$ at small scales, manifesting the Magneto-Archimedean-Coriolis (MAC) balance \citep[see, e.g.,][]{davidson2013,augustson2019,schwaiger2021}.

Assuming the CIA and MAC balances, the following scaling relations can be derived in terms of the flux Coriolis number Co$_*$ (see Appendix~\ref{appendix:scaling} for detailed derivations)
\begin{eqnarray}
&& \mathrm{Length \ scale:} \ \ \    \ell \propto \left\{
  \begin{array}{ll}
    \mathrm{Co_*}^{-3/5} & \ (\mathrm{CIA}) \\
    \mathrm{Co_*}^{-1/3} & \ (\mathrm{MAC}),
  \end{array}
  \right. \label{eq:scaling_length} \\
&& \mathrm{Velocity:} \ \ \    v \propto \left\{
  \begin{array}{ll}
    \mathrm{Co_*}^{-1/5} & \ (\mathrm{CIA}) \\
    \mathrm{Co_*}^{-1/3} & \ (\mathrm{MAC}),
  \end{array}
  \right. \label{eq:scaling_vrms}   \\
&& \mathrm{Entropy \ fluctuation:} \ \ \    s \propto \left\{
  \begin{array}{ll}
    \mathrm{Co_*}^{1/5} & \ (\mathrm{CIA}) \\
    \mathrm{Co_*}^{1/3} & \ (\mathrm{MAC}).
  \end{array}
  \right. \label{eq:scaling_s1rms}    
\end{eqnarray}
In Figs.~\ref{fig:Ekin_kmean}c, \ref{fig:vrmsBrms}e, and \ref{fig:s1rms}e, our simulated Co$_*$ dependences are compared with these theoretical scalings.
It is clearly seen that our results are consistent with the predictions by CIA and MAC balances at Co$_* \ge 5$.

One might question the robustness of the above results, particularly when the numerical resolution is further increased. 
Hints are provided by \citet{kapyla2024} who conducted HD simulations at varying Re and demonstrated that the scaling of convective velocity $v_{\rm rms}$ with Coriolis number remains unaffected, although the overall amplitude of $v_{\rm rms}$ increases with Re. 
This is because viscous force does not influence the dominant force balance. 
Given that the CIA and MAC balances are well established in our simulations, the above scaling relations at high-Co$_*$ regime are expected to remain robust even at higher resolutions, although the absolute values of $v_{\rm rms}$ (or $s_{\rm rms}$) and the degree of suppression (enhancement) from HD to MHD will be affected.


\subsection{Dominant scale of columnar convection} \label{sec:invcascade}

\begin{figure*}    
\begin{center}
\includegraphics[width=0.9\linewidth]{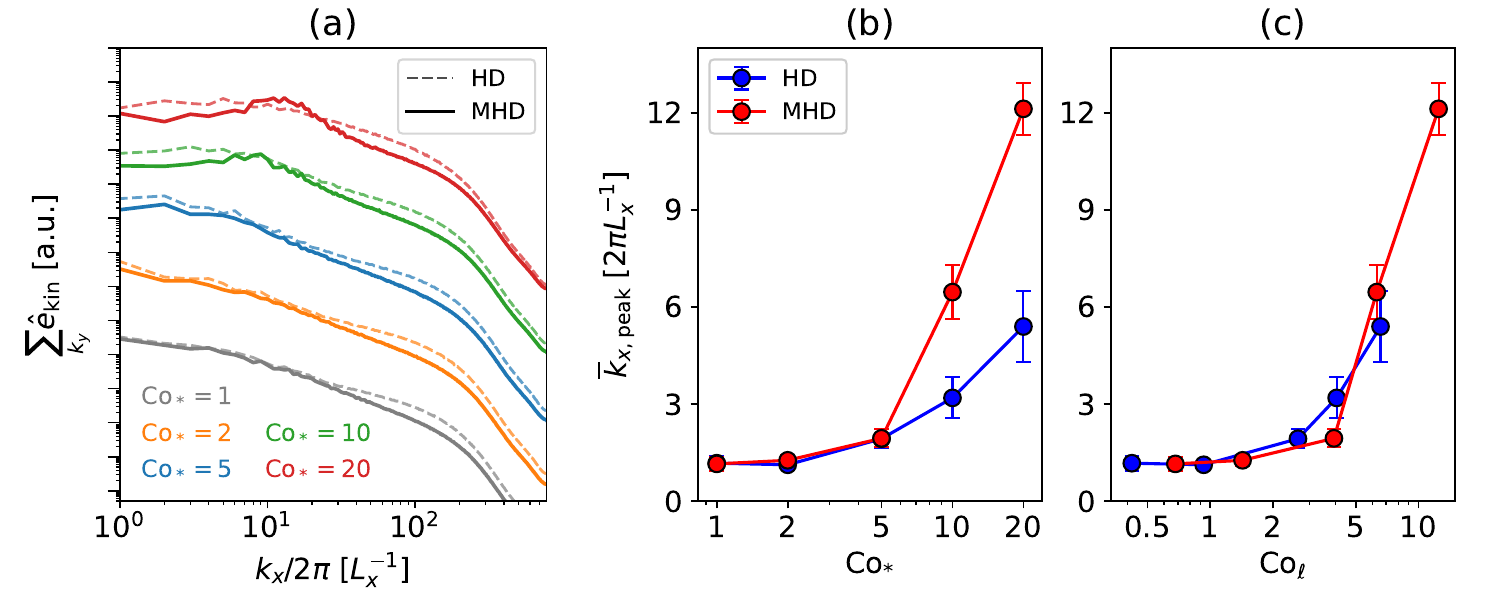}
\caption{
(a) Normalized kinetic energy spectra, $\sum_{k_y} \hat{e}_{\rm kin}$, in the middle convection zone ($z=1.1~H_r$).
Different colors represent different Co$_*$. Dashed and solid lines show the results from HD and MHD simulations, respectively.
(b) Longitudinal wavenumbers where the kinetic energy spectra peak, $k_{x,\rm peak}$, plotted against the flux Coriolis number Co$_*$.
Blue and Red circles are the results for HD and MHD cases.
(c) Same as panel b but as a function of the dynamical Coriolis number Co$_\ell =2\Omega_0 \overline{\ell}_x / \overline{v}_{\rm rms}$.
}
\label{fig:Ekin_ky0}
\end{center}
\end{figure*}

In this subsection, we discuss how SSD affects the dominant scale of columnar convective modes.
From Figs.~\ref{fig:snap_vz_Co20}a and b, it is seen that the SSD not only makes the columnar patterns more self-organized but also shifts the dominant longitudinal scale of convection narrower.
To quantify the difference in the length scales of columnar convection, we compare in Fig.~\ref{fig:Ekin_ky0}a the kinetic energy spectra between HD and MHD cases in the middle convection zone ($z=1.1~H_r$).
Figure~\ref{fig:Ekin_ky0}b shows the peak longitudinal wavenumber $\overline{k}_{x,\rm peak}$ as a function of Co$_*$.
In both HD and MHD cases, there is a general trend that $\overline{k}_{x,\rm peak}$ increases with Co$_*$.
However, it is obvious that the MHD simulations show a much more rapid increase of $\overline{k}_{x,\rm peak}$ with Co$_*$ compared to the HD counterparts.
As a consequence, at high-Co$_*$ regime, the convective columns have a smaller length scale in longitude with the presence of the SSD.
This behavior cannot be explained by the CIA/MAC scaling relations (Eq.~\ref{eq:scaling_length}).

We find that, through the suppression in the convective velocity $v_{\rm rms}$ and the increase in the (mean) convective length scale $\ell_{x} = 2\pi/k_{x,\rm mean}$, the SSD can change the effective rotational influence in the simulations.
This is typically measured by the \textit{dynamical} Coriolis number 
\begin{eqnarray}
    && \mathrm{Co}_{\ell} = \frac{2\Omega_0 \overline{\ell}_x}{\overline{v}_{\rm rms}}.
\end{eqnarray}
As reported in Table~\ref{table:1}, Co$_\ell$ is systematically larger in MHD than in HD, implying that the MHD simulations are operating in more rotationally constrained regimes compared to the HD counterparts.
Consequently, the establishment of the columnar patterns is more enhanced and the longitudinal size of these columns becomes smaller in our MHD simulations.
In fact, we find that the peak wavenumbers $\overline{k}_{x,\rm peak}$ in both HD and MHD cases manifest very similar dependences on the dynamical Coriolis number Co$_\ell$ (Fig.~\ref{fig:Ekin_ky0}c).
This provides supporting evidence that the dominant scale of columnar convection is largely determined by Co$_\ell$, a parameter that is sensitively affected by the SSD.

We also note that there are other underlying mechanisms that can potentially explain the difference in $\overline{k}_{x,\rm peak}$ between HD and MHD.
First possibility is the direct Lorentz force feedback. As will be shown in \S~\ref{sec:Maxwell}, the large-scale shear motions in the columnar convection can generate the Maxwell stress to redistribute the angular momentum in the convection zone. As a back-reaction, the kinetic energy of columnar convection is suppressed at large-scales, which can also cause a shift of $\overline{k}_{x,\rm peak}$ towards higher wavenumbers.
Another possible mechanism comes through the change in the mean background stratification.
When the SSD is present, the buoyant driving can be selectively suppressed at large scales due to the reduced superadiabaticity and the existence of the weakly subadiabatic layer near the base of the convection zone (Fig.~\ref{fig:delta}).
Yet, it is not easy to disentangle these mechanisms and to pinpoint the primary cause. Such an analysis is beyond the scope of this paper.


\subsection{Mean flow generation} \label{sec:meanflows}

\begin{figure*}
\begin{center}
\includegraphics[width=0.8\linewidth]{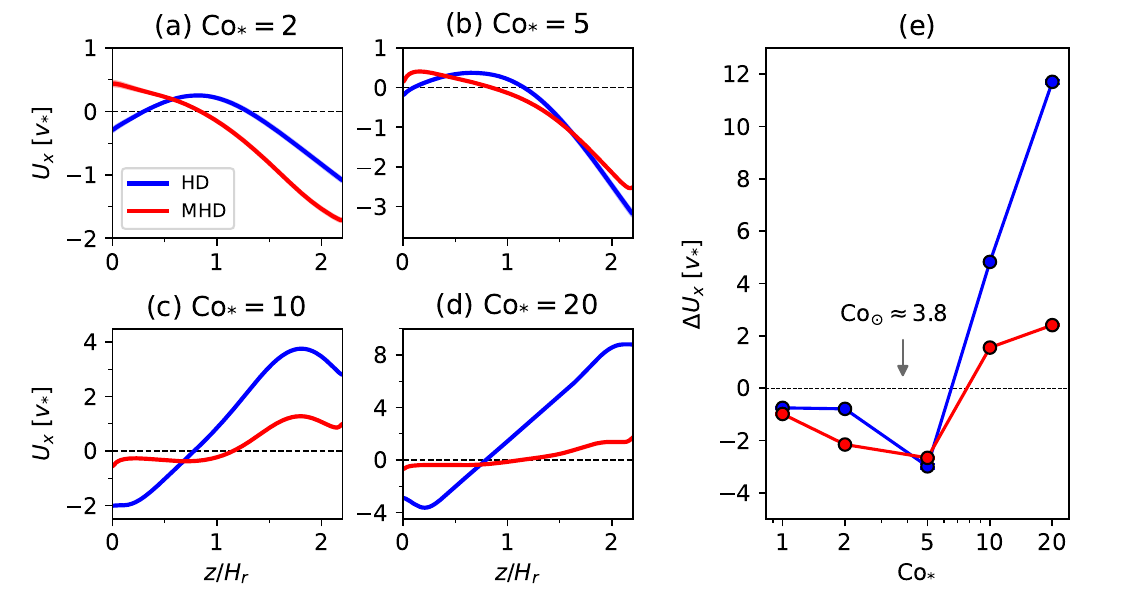}
\caption{
Profiles of the mean flows $U_{x}(z)$.
(a--d) Vertical profiles of $U_{x}(z)$ for different $\mathrm{Co}_{*}$.
Blue and red colors represent the results from the HD and MHD simulations.
(e) The vertical shear amplitudes of the mean flow, $\Delta U_{x}=U_{x}(z_{\mathrm{max}})-U_{x}(z_{\mathrm{min}})$, as a function of $\mathrm{Co}_{*}$.
}
\label{fig:mean_vx}
\end{center}
\end{figure*}
\begin{figure*}
\begin{center}
\includegraphics[width=0.8\linewidth]{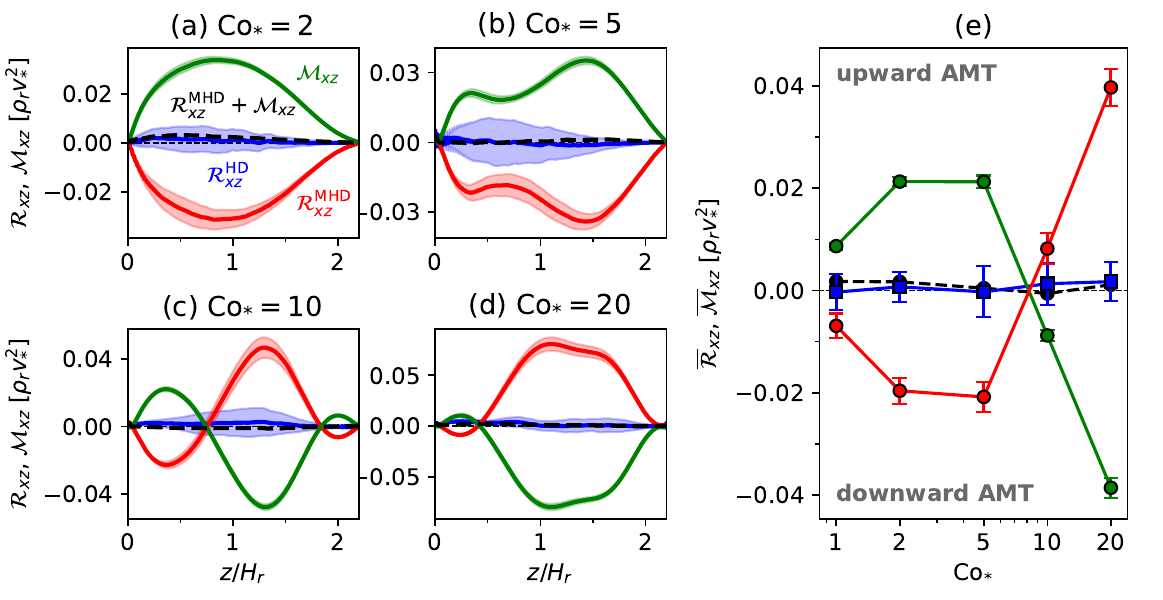}
\caption{
Profiles of the Reynolds and Maxwell stresses, defined by Eq.~(\ref{eq:RSMS}).
Positive and negative values correspond to the upward and downward AMT.
(a--d) Vertical profiles of the Reynolds stress $\mathcal{R}_{xz}$ for different values of Co$_{*}$.
Blue and red colors represent the results from the HD and MHD cases.
Green curves show the vertical profiles of the Maxwell stress $\mathcal{M}_{xz}$ and the black dashed curves indicate the sum $\mathcal{R}_{xz}+\mathcal{M}_{xz}$ in the MHD simulations.
(e) The volume-averaged Reynolds and Maxwell stresses, $\overline{\mathcal{R}}_{xz}$ and $\overline{\mathcal{M}}_{xz}$, plotted as functions of Co$_{*}$.
}
\label{fig:RSMS}
\end{center}
\end{figure*}
\begin{figure*}
\begin{center}
\includegraphics[width=0.9\linewidth]{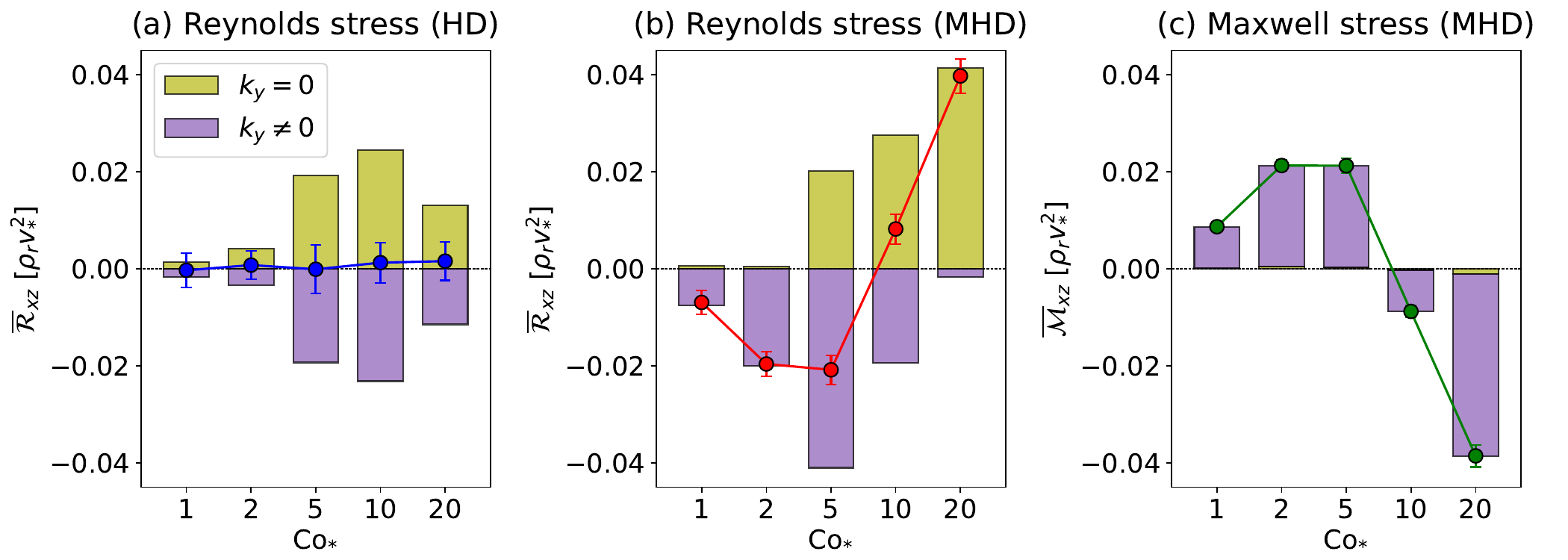}
\caption{
Breakdown of the contributions from the $k_y=0$ and $k_y \ne 0$ components to the Reynolds and Maxwell stresses.
(a,b) Volume-averaged Reynolds stress $\overline{\mathcal{R}}_{xz}$ as a function of Co$_{*}$ from the HD and MHD simulations.
Yellow and purple colors correspond to $\overline{\mathcal{R}}_{xz}^{k_y=0}$ and $\overline{\mathcal{R}}_{xz}^{k_y \ne 0}$, respectively.
(c) Volume-averaged Maxwell stress $\overline{\mathcal{M}}_{xz}$ from MHD simulations.
}
\label{fig:RSMS_ky0}
\end{center}
\end{figure*}
\begin{figure*}
\begin{center}
\includegraphics[width=0.99\linewidth]{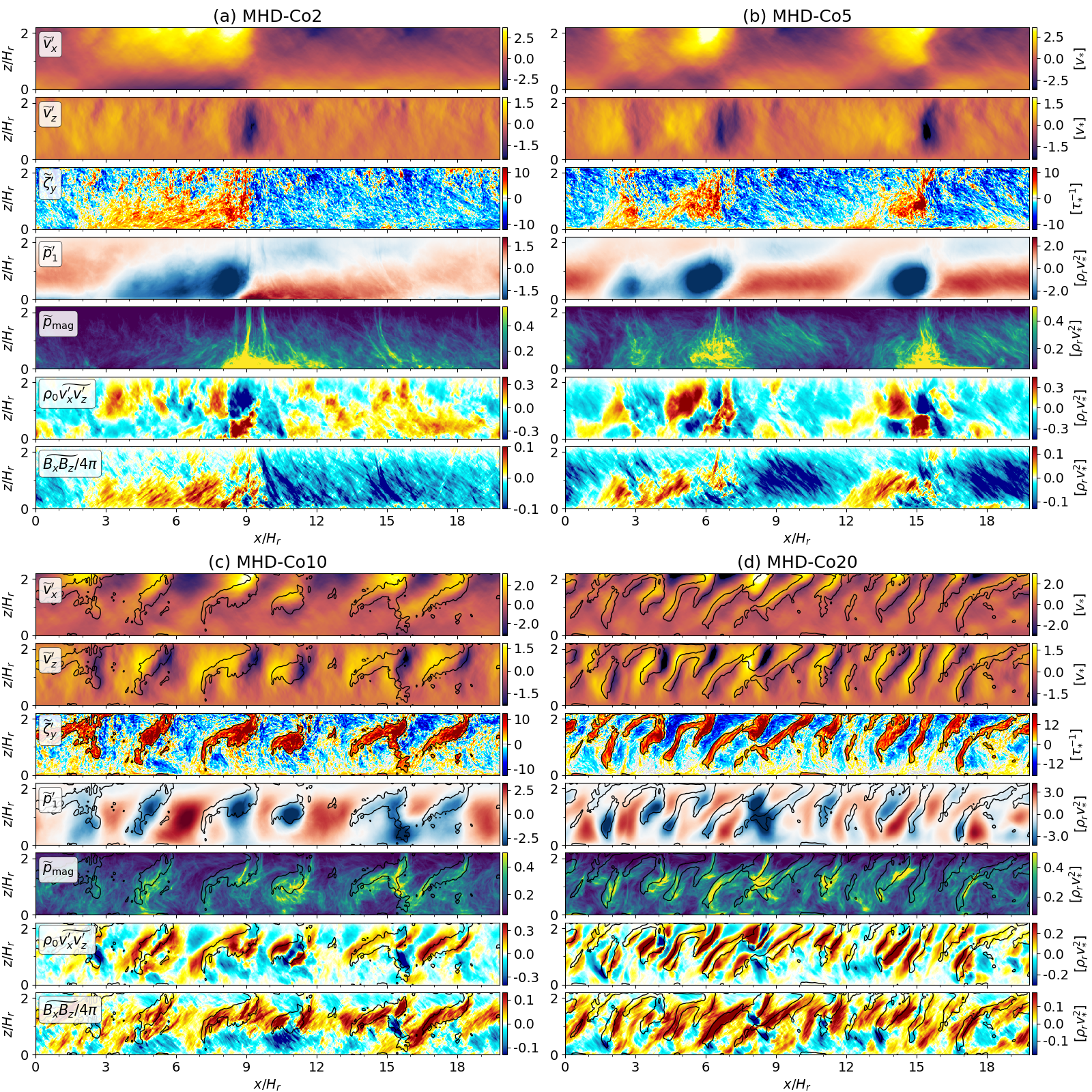}
\caption{
Snapshots of the $y$-averaged quantities from MHD simulations with Co$_{*}=2, \ 5, \ 10$, and $20$.
Each panel, from top to bottom, displays the $y$-averaged profiles of $x$-velocity $\widetilde{v}'_{x}$, $z$-velocity $\widetilde{v}'_{z}$, $y$-vorticity $\widetilde{\zeta}'_{y}$, pressure perturbation $\widetilde{p}'_{1}$, magnetic energy density $\widetilde{p}_{\rm mag}$, velocity correlation $\rho_0 \widetilde{v_x^{\prime} v_z^{\prime}}$, and magnetic field correlation $\widetilde{B_x B_z}/4\pi$.
In rapidly rotating cases MHD-Co10 and MHD-Co20 (panels c--d), black curves indicate contour lines of the $y$-vorticity at $\widetilde{\zeta}'_{y}=2 \ \tau_{*}^{-1}$.
}
\label{fig:side_yave_MHD}
\end{center}
\end{figure*}
\begin{figure*}
\begin{center}
\includegraphics[width=0.875\linewidth]{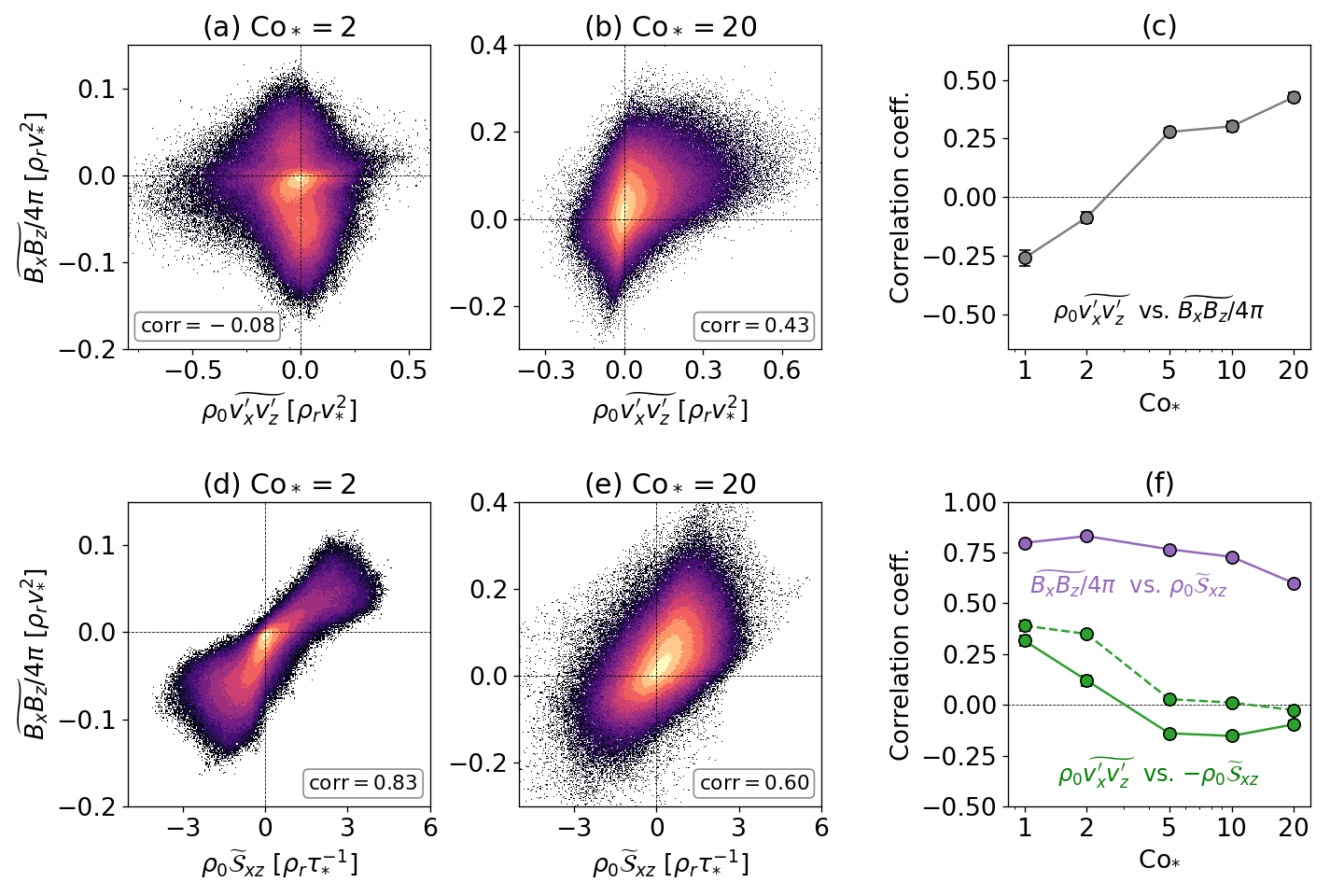}
\caption{
(Top) Correlations between the Maxwell stress $\widetilde{B_x B_z} / 4\pi$ and the Reynolds stress $\rho_0 \widetilde{v'_x v'_z}$ and (bottom) correlations between $\widetilde{B_x B_z} / 4\pi$ and the velocity strain tensor $\rho_0 \widetilde{\mathcal{S}}_{xz}= \rho_0 (\partial_z \widetilde{v}_x + \partial_{x} \widetilde{v}_z)$ in MHD simulations.
(a,b) Probability density functions (PDFs) between $\widetilde{B_x B_z} / 4\pi$ and $\rho_0 \widetilde{v'_x v'_z}$ in MHD-Co2 and MHD-Co20 cases.
(c) Correlation coefficient as a function of Co$_*$.
(d,e) PDFs between $\widetilde{B_x B_z} / 4\pi$ and $\rho_0 \widetilde{\mathcal{S}}_{xz}$ in MHD-Co2 and MHD-Co20 cases.
(f) Correlation coefficients $\widetilde{B_x B_z} / 4\pi$ vs. $\rho_0 \widetilde{\mathcal{S}}_{xz}$ (purple) and $\rho_0 \widetilde{v'_x v'_z}$ vs. $-\rho_0 \widetilde{\mathcal{S}}_{xz}$ (green) as functions of Co$_*$.
Green dashed and solid lines show the results from HD and MHD simulations, respectively.
}
\label{fig:PDF_MS_all}
\end{center}
\end{figure*}

In our simulations, large-scale mean flows are generated only in the longitudinal direction, $U_{x}(z)=\langle v_{x} \rangle$, corresponding to the radial differential rotation at the equator.
Figures~\ref{fig:mean_vx}a--d show profiles of $U_{x}(z)$ established in our HD and MHD simulations for different values of Co$_{*}$.
A mean flow regime can be measured by the vertical shear $\Delta U_x = U_x(z_{\rm max}) - U_x(z_{\rm min})$, as shown in Fig.~\ref{fig:mean_vx}e.
For Co$_{*}=1, \ 2$, and $5$, the mean flows have negative vertical shear $\Delta U_x <0$ with a faster base and a slower surface.
This might correspond to anti-solar differential rotation profile in a spherical shell geometry.
On the other hand, for Co$_{*}=10$ and $20$, the mean flows show a positive vertical shear $\Delta U_x >0$ with a faster surface and a slower base, likely corresponding to a solar-like rotational profile in the spherical case.
We find that an inclusion of the SSD has a primary impact of quenching the mean flow amplitudes at high Co$_*$, as reported in the literature \citep[e.g.,][]{brun2004,kapyla2017b,matilsky2022,warnecke2025}.

Another important effect of the SSD on the mean flow $U_x$ is the excitation of torsional oscillations, i.e., temporal variations of the mean flow. In our model, these oscillations arise as a by-product of the establishment of the MAC balance, where torsional Alfvén waves propagate vertically as disturbances to the so-called Taylor state \citep[][]{taylor1963,braginsky1970,wicht2010}. Our analysis indicates that their amplitudes are too small (few percent of the rms velocity fluctuation) to have a dynamical impact in the simulation. Therefore, a more detailed discussion is deferred to the Appendix~\ref{appendix:TO}.

We now discuss the generation mechanism of the mean flows.
Taking a horizontal average of the $x$-component of Eq.~(\ref{eq:motion}), we have
\begin{eqnarray}
    && \frac{\partial U_x}{\partial t} = -\frac{\partial}{\partial z} \left( \mathcal{R}_{xz}+\mathcal{M}_{xz} \right), \label{eq:dUxdt}
\end{eqnarray}
where
\begin{eqnarray}
    &&  \mathcal{R}_{xz}=  \rho_0 \langle v_x' v_z' \rangle , \ \ \ \ 
     \mathcal{M}_{xz} = - \frac{\langle B_x B_z \rangle}{4\pi}, \label{eq:RSMS}
\end{eqnarray}
are the turbulent Reynolds and Maxwell stresses, respectively.
Positive and negative values of $\mathcal{R}_{xz}$ and $\mathcal{M}_{xz}$ correspond to vertically upward and downward fluxes of the $x$ (angular) momentum.
Figures~\ref{fig:RSMS}a--d show vertical profiles of $\mathcal{R}_{xz}$ and $\mathcal{M}_{xz}$ for different Co$_*$.
Their height-averaged values are shown as functions of Co$_*$ in Fig.~\ref{fig:RSMS}e.
Note that, in HD, $\mathcal{R}_{xz}$ must vanish in a statistically stationary state ($\partial U_x / \partial t \approx 0$).
This reflects a balance between the diffusive and non-diffusive parts (the $\Lambda$ effect) of the Reynolds stress $\mathcal{R}_{xz}$, which transport angular momentum in opposite directions \citep[][]{kitchatinov2013,barekat2021}.
In MHD, $\mathcal{R}_{xz}$ and $\mathcal{M}_{xz}$ must balance with each other such that their sum vanishes statistically.
At low Co$_* \  (\le 5)$ where the mean flow is anti-solar ($\Delta U_x <0$), the Reynolds stress is negative ($\overline{\mathcal{R}}_{xz} <0)$ while the Maxwell stress is positive ($\overline{\mathcal{M}}_{xz} >0$).
Conversely, at high Co$_* \ (\ge 10)$ where the mean flow is solar-like ($\Delta U_x > 0$), the signs of the Reynolds and Maxwell stresses are flipped; $\overline{\mathcal{R}}_{xz}>0$ and $\overline{\mathcal{M}}_{xz}<0$.
Thus, we conclude that the mean flows are primarily generated by the angular momentum transport (AMT) through the Reynolds stress $\mathcal{R}_{xz}$ with the Maxwell stress $\mathcal{M}_{xz}$ only playing a suppressive role.
At low Co$_*$, the anti-solar mean flows are established by the downward AMT by weakly rotationally constrained convection \citep[][]{foukal1975,guerrero2013,featherstone2015}, whereas at high Co$_*$, the upward AMT by rotationally constrained convective columns leads to the solar-like mean flows \citep[][]{hotta2014b,matilsky2019,bekki2022b} (see also Appendix~\ref{appendix:origin_tilt} for a schematic explanation).

To assess the impact of the columnar convective modes on AMT, we decompose the Reynolds stress into contributions from the $y$-averaged flows (representing the $k_y=0$ columnar modes) $\mathcal{R}_{xz}^{k_y=0}$ and those from the velocity fluctuations $\mathcal{R}_{xz}^{k_y \ne 0}$ as
\begin{eqnarray}
 &&   \mathcal{R}_{xz} = \underbrace{\rho_0 \langle \widetilde{v}_x \widetilde{v}_z \rangle }_{\mathcal{R}_{xz}^{k_y=0}}
    + \underbrace{\rho_0 \langle v^{\prime(y)}_x v^{\prime(y)}_z \rangle}_{\mathcal{R}_{xz}^{k_y \ne 0}}.
\end{eqnarray}
Figures~\ref{fig:RSMS_ky0}a and b show the volume-averaged Reynolds stress contributions $\overline{\mathcal{R}}_{xz}^{k_y=0}$ and $\overline{\mathcal{R}}_{xz}^{k_y \ne 0}$.
There is a general tendency for the $k_y=0$ columnar modes to transport angular momentum upward ($\overline{\mathcal{R}}_{xz}^{k_y=0}>0$) and the small-scale fluctuating velocities transport angular momentum downward ($\overline{\mathcal{R}}_{xz}^{k_y \ne 0}<0$).
In HD cases, these two contributions cancel each other.
In particular, at high Co$_*$, $\overline{\mathcal{R}}_{xz}^{k_y=0}>0$ corresponds to a driving of the mean flow $\Delta U_x >0$ and $\overline{\mathcal{R}}_{xz}^{k_y \ne 0}<0$ corresponds to a turbulent diffusion.
These tendencies are consistent with scale-by-scale analyses by \citet{mori2023a} and \citet{kapyla2023}, who found downward AMT from small-scale turbulence and upward AMT from large-scale columnar modes near the equator. We note, however, that their analyses are based on longitudinal scale decompositions, whereas we distinguish the $k_y=0$ columnar modes from the $k_y \neq 0$ fluctuations; therefore a one-to-one correspondence should not be assumed.
In MHD cases, the cancellation between $\overline{\mathcal{R}}_{xz}^{k_y=0}$ and $\overline{\mathcal{R}}_{xz}^{k_y \ne 0}$ breaks down.
At low Co$_*$, the $k_y \ne 0$ convective motions that are weakly influenced by rotation dominantly transport angular momentum downward.
As Co$_*$ increases, the rotationally constrained columnar modes become increasingly dominant, resulting in net upward AMT (see Appendix~\ref{appendix:origin_tilt}).
Here, the downward AMT by $k_y \ne 0$ flows is substantially reduced because the effective eddy diffusion due to small-scale turbulence is suppressed by the SSD Lorentz force.
The resulting net downward (upward) angular momentum flux at low (high) Co$_*$ is then compensated by the upward (downward) flux of the Maxwell stress where all the contributions are from $k_y \ne 0$ magnetic fields (Fig.~\ref{fig:RSMS_ky0}c).

\subsection{Origin of Maxwell stress: Interplay with convective columns} \label{sec:Maxwell}

In this subsection, we discuss the generation mechanisms of the Maxwell stress $\mathcal{M}_{xz}$.
To this end, we compare the spatial distribution of the magnetic field correlation $\widetilde{B_x B_z}$ (averaged over $y$) with various $y$-averaged quantities, as shown in Fig.~\ref{fig:side_yave_MHD}.
Let us first focus on the low-Co$_*$ regime where the Maxwell stress transports angular momentum upward ($\overline{\mathcal{M}}_{xz}>0$).
It is seen from Figs.~\ref{fig:side_yave_MHD}a and b that the magnetic field correlation $\widetilde{B_x B_z}$ is not necessarily generated in downflow regions where strong magnetic energy is concentrated and where the velocity correlation $\widetilde{v'_x v'_z}$ is dominantly generated.
As a result, the spatial distribution of $\widetilde{B_x B_z}$ is poorly correlated with that of $\widetilde{v'_x v'_z}$ (Fig.~\ref{fig:PDF_MS_all}a).
This implies that the generation of the Maxwell stress is not a consequence of a parallelization of the small-scale magnetic fields with the small-scale flows, as proposed by \citet{hotta2022}.
Rather, in our simulations, spatial distribution of $\widetilde{B_x B_z}$ is very well correlated with those of the $y$-vorticity $\widetilde{\zeta}'_y$ and also of the pressure perturbation $\widetilde{p}'_1$. 
The latter is in fact a side effect required by the geostrophic balance (anti-correlation between $\widetilde{\zeta}'_y$ and $\widetilde{p}'_1$). 

A good correlation between spatial distributions of $\widetilde{B_x B_z}$ and $\widetilde{\zeta}'_y$ suggests that the shear in the columnar convection is the main origin of the Maxwell stress in our simulations. 
From the induction Eq.~(\ref{eq:induction}), we obtain
\begin{eqnarray}
     && \frac{\partial}{\partial t} \widetilde{B_x B_z} = [...] + \widetilde{B_x^2} \frac{\partial \widetilde{v}'_z}{\partial x} + \widetilde{B_z^2} \frac{\partial \widetilde{v}'_x}{\partial z},
\end{eqnarray}
indicating that $\widetilde{B_x B_z} >0 \ (<0)$ can be generated by shear motions in the $y$-averaged flows, $\partial \widetilde{v}'_x /\partial z >0 \ (<0)$ and $\partial \widetilde{v}'_z /\partial x >0 \ (<0)$.
Therefore, if the main generation mechanism of the Maxwell stress is the columnar shear, $\widetilde{B_x B_z}$ must be strongly correlated with the velocity strain tensor $\widetilde{\mathcal{S}}_{xz}$ defined as
\begin{eqnarray}
    && \widetilde{\mathcal{S}}_{xz} =  \frac{\partial \widetilde{v}'_x}{\partial z} + \frac{\partial \widetilde{v}'_z}{\partial x} . \label{eq:Seddy}
\end{eqnarray}
This is clearly shown to be the case from Figs.~\ref{fig:PDF_MS_all}d--f.
Since the longitudinal separation between columnar downflows is large at low Co$_*$, the vertical shear of the zonal velocity dominates over the zonal shear of the vertical flow, $|\partial \widetilde{v}'_x /\partial z| \gg |\partial \widetilde{v}'_z /\partial x|$.
In such a case, both $\widetilde{\mathcal{S}}_{xz}$ and $\widetilde{\zeta}'_y$ are primarily determined by $\partial \widetilde{v}'_x /\partial z$, which explains why $\widetilde{B_x B_z}$ and $\widetilde{\zeta}'_y$ are so well correlated in Fig.~\ref{fig:side_yave_MHD}a and b.
When averaged over the whole volume, the net correlation $\langle B_x B_z \rangle$ becomes negative because the $y$-vorticity consists of spatially localized $\widetilde{\zeta}_y >0$ and broadened $\widetilde{\zeta}'_y <0$ due to the nonlinear effect (see Appendix~\ref{appendix:origin_asymmetry} for its physical explanation).

Next, we discuss the results in high-Co$_*$ regime (Figs.~\ref{fig:side_yave_MHD}c and d).
Under strong rotational influences, the longitudinal separation of the convective columns decreases and the zonal shear of the vertical flow becomes non-negligible.
Consequently, $\widetilde{B_x B_z}$ is no longer well correlated with $\widetilde{\zeta}'_y$.
Instead, $\widetilde{B_x B_z}$ is now positively correlated with $\widetilde{v_x' v_z'}$ (Figs.~\ref{fig:PDF_MS_all}b and c).
This is not surprising because $\widetilde{v_x' v_z'}$ and $\widetilde{B_x B_z}$ are both rooted in the $y$-coherent columnar pattern.
We note, however, that this positive correlation between $\widetilde{v_x' v_z'}$ and $\widetilde{B_x B_z}$ is not likely due to the local alignment of small-scale velocity to small-scale magnetic fields.
As clearly manifested by strong correlation between $\widetilde{B_x B_z}$ and  $\widetilde{\mathcal{S}}_{xz}$ in Fig.~\ref{fig:PDF_MS_all}e, a main generation mechanism of a positive magnetic field correlation $\widetilde{B_x B_z} >0$ remains the shear motions in the columnar convection (both $\partial \widetilde{v}'_x /\partial z$ and $\partial \widetilde{v}'_z /\partial x$).
On the other hand, a positive velocity correlation $\widetilde{v_x' v_z'} >0$ is established as a consequence of the tilted pattern of the $y$-averaged columns (see Appendix~\ref{appendix:origin_tilt}).

Our simulations clearly show that the Maxwell stress is primarily generated by the shear in the columnar motions, regardless of the rotational regime. This is in contrast to the results of \citet{hotta2022}, who argue that the Maxwell stress originates mainly from the Reynolds stress. One possible reason for this discrepancy lies in the differences in the simulation setups: their model covers a full spherical shell, in contrast to our equatorial local model, and the interaction between SSD and rotating convection at middle to high latitudes (where the columnar convection is absent) may play a dominant role. Moreover, in their simulations, the presence of super-equipartition magnetic fields can significantly suppress the velocity shear, which may inhibit the generation of Maxwell stress through shear.

\subsection{Amplitudes of columnar modes} \label{sec:amplitudes}

\begin{figure}
\begin{center}
\includegraphics[width=0.85\linewidth]{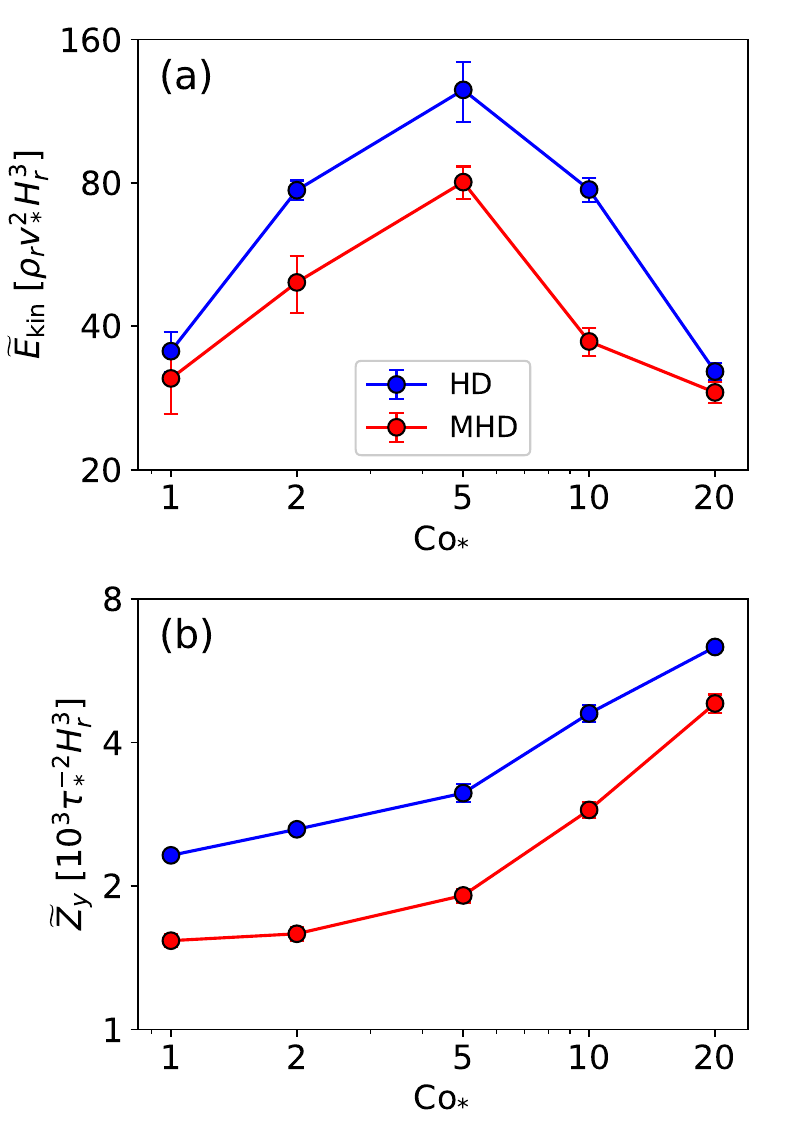}
\caption{
Volume-integrated (a) kinetic energy $\widetilde{E}_{\rm kin}$ and (b) $y$-enstrophy $\widetilde{Z}_y$ of the $y$-averaged flow as functions of Co$_*$. 
The definitions of $\widetilde{E}_{\rm kin}$ and  $\widetilde{Z}_y$ are given by Eqs.~(\ref{eq:Ekin_ky0}) and (\ref{eq:Zy_ky0}).
}
\label{fig:EkinEnst_ky0_vol}
\end{center}
\end{figure}

In all our simulations, we find that the velocity strain tensor $\widetilde{\mathcal{S}}_{xz}$ is not correlated with $\widetilde{v_x' v_z'}$ but is closely correlated with $\widetilde{B_x B_z}$ (Fig.~\ref{fig:PDF_MS_all}f).
This suggests that the Lorentz force associated with $\widetilde{B_x B_z}$ plays a significant role as an effective viscosity on the columnar convection, leading to a suppression of the columnar convective modes.
Figure~\ref{fig:EkinEnst_ky0_vol}a shows the volume-integrated kinetic energy of the $y$-averaged flows
\begin{eqnarray}
    && \widetilde{E}_{\rm kin} = \int_V \frac{\rho_0}{2} (\widetilde{v'}_x^{2} + \widetilde{v'}_z^2) dV, \label{eq:Ekin_ky0}
\end{eqnarray}
as a function of Co$_*$.
There is a general tendency seen in both HD and MHD cases that $\widetilde{E}_{\rm kin}$ first increases and then decreases as Co$_*$ increases.
The initial increase in $\widetilde{E}_{\rm kin}$ is simply due to the establishment of a coherent columnar pattern.
The decrease in $\widetilde{E}_{\rm kin}$ at Co$_* >5$ can be attributed to the suppression of convective velocity in a strongly rotationally constrained regime (Eq.~\ref{eq:scaling_vrms}).
It would also be instructive to use the volume-integrated $y$-enstrophy 
\begin{eqnarray}
    && \widetilde{Z}_y = \int_V \widetilde{\zeta'}_y^2 dV, \label{eq:Zy_ky0}
\end{eqnarray}
as a measure of the amplitudes of the columnar modes.
As shown in Fig.~\ref{fig:EkinEnst_ky0_vol}b, $\widetilde{Z}_y$ increases monotonically with Co$_*$ in both HD and MHD cases.
Suppression of the columnar convection by the SSD can be seen in both $\widetilde{E}_{\rm kin}$ and in $\widetilde{Z}_y$.
The suppression is the most drastic at intermediate regime Co$_* =5$ where both $\widetilde{E}_{\rm kin}$ and $\widetilde{Z}_y$ are reduced by about $40$\% from HD to MHD.


\section{Summary and conclusions} \label{sec:summary}

In this paper, we carried out a series of hydrodynamic (HD) and magnetohydrodynamic (MHD) simulations of highly turbulent rotating compressible convection using a local $f$-plane box model at the equator. This model setup enables us to focus on the effects of small-scale dynamo (SSD) on rotating columnar convection. We performed a parameter study in which the rotation rate was varied while fixing the energy flux, covering the flux Coriolis number Co$_*$ from 1 to 20.

We find that the SSD has a significant impact on the properties of convective heat transport. As reported in previous non-rotating studies, the convective velocity $v_{\rm rms}$ tends to be suppressed due to the Lorentz force feedback \citep[][]{rempel2014,hotta2015}. Under strong rotational influences, we find that $v_{\rm rms}$ decreases with Co$_*$ more rapidly in MHD cases than in their HD counterparts. That is, convection tends to be suppressed much more significantly by rotation when the SSD exists. This rapid decrease of convective speed $v_{\rm rms}$ with Co$_*$ is compensated by a rapid increase in the entropy fluctuations $s_{\rm rms}$ with increasing Co$_*$, which is required for rotating convection to transport the fixed amount of enthalpy upward. One of the most striking outcomes of this entropy enhancement by the SSD is the existence of a weakly subadiabatic layer near the base of the convection zone, which persists even in a strongly rotationally constrained regime (Co$_* = 20$). This is in significant contrast to rotating HD convection, where the weakly subadiabatic layer vanishes when the rotational influence becomes strong enough \citep[e.g.,][]{kapyla2024}.
The potential existence of this subadiabatic layer in rapidly-rotating stars has profound implications for dynamo processes in young Sun-like stars \citep[e.g.,][]{brown2011,viviani2018,kapyla2019}.

These results can be understood by investigating how SSD alters the dominant force balance. Our analysis reveals that, at small scales, the pressure gradient force is balanced by the Lorentz force rather than by inertia when the SSD is present, establishing the so-called magnetostrophic balance. On the other hand, at large scales, a geostrophic balance is always maintained. Consequently, the SSD causes a transformation of the regime of rotating convection from a quasi-geostrophic (QG) Coriolis-Inertia-Archimedean (CIA) balance to a QG Magnetic-Archimedean-Coriolis (MAC) balance. In fact, it is shown that the Co$_*$ dependences of $v_{\rm rms}$ and $s_{\rm rms}$ in our HD and MHD simulations are consistent with the theoretical scaling relations of the CIA and MAC balances in the rapidly rotating limit (Co$_* \gtrsim 5$).

In rotating columnar convection, there are two distinct characteristic length scales. When a typical convective length scale is measured by a mean longitudinal wavenumber $\overline{k}_{x,\rm mean}$ (weighted by the kinetic energy spectrum), we find that $\overline{k}_{x,\rm mean}$ becomes smaller in MHD cases, i.e., the typical size of convective cells becomes larger, because the kinetic energy is suppressed at small scales due to the SSD Lorentz force feedback. In the rapidly rotating regime (Co$_* \gtrsim 5$), $\overline{k}_{x,\rm mean}$ increases with Co$_*$ in both HD and MHD cases, roughly consistent with the CIA and MAC scalings, respectively. 
On the other hand, the peak wavenumber $\overline{k}_{x,\rm peak}$ (at which the kinetic energy spectrum has a local maximum) represents the dominant scale of the prograde-propagating columnar convective (thermal Rossby) modes, and it does not follow the CIA/MAC scalings.
Our analysis reveals that the SSD enhances the effective rotational influence on columnar convection by increasing the dynamical Coriolis number Co$_\ell \propto 1/(\overline{v}_{\rm rms} \overline{k}_{x,\rm mean})$.
As a result, the dominant scale of columnar convection becomes smaller when the SSD is present.

In addition to the convective heat transport, we also find that the SSD has a significant impact on the convective angular momentum transport (AMT). As already reported in previous studies \citep[e.g.,][]{gastine2013,guerrero2013,mabuchi2015,featherstone2015}, we see a transition from the "anti-solar" mean flow profile (with a faster base and a slower surface) to the "solar-like" mean flow profile (with a faster surface and a slower base) as Co$_*$ increases. We find that the SSD acts to suppress the amplitudes of the mean shear flows, as it provides a Maxwell stress that counteracts the Reynolds stress. This result is consistent with many previous studies \citep[][]{brun2004,nelson2013,augustson2015,kapyla2017b,warnecke2025}. 
Some spherical dynamo simulations \citep[e.g.,][]{fan2014,karak2015,hotta2022,soderlund2025} further reported that the magnetic field can change the rotational profile from anti-solar to solar-like near the solar parameter regime (Co$_* \approx 2-5$).
Such a drastic transition was not seen in this study.
This can be primarily attributed to the lack of AMT via meridional circulation - an essential ingredient for the generation of differential rotation in spherical models - in our simplified $f$-plane box model.

In \citet{hotta2022}'s simulation, a key to achieving solar-like differential rotation was the radially-outward AMT by the SSD Maxwell stress. They proposed that this Maxwell stress arises as a back-reaction to the Reynolds stress because the small-scale magnetic fields tend to be parallel to the small-scale flows. In our simulations, we do not find clear evidence for this so-called \textit{Punching-ball} effect. In fact, we find that the $y$-averaged profile of the Maxwell stress $\widetilde{B_x B_z}$ is poorly correlated with that of the Reynolds stress $\widetilde{v_x' v_z'}$ but is well correlated with those of the $y$-vorticity $\widetilde{\zeta}'_y$ or the vertical velocity shear $\partial \widetilde{v}'_x/\partial z$ of the columnar convection. This implies that the Maxwell stress in our simulations is mainly generated by the velocity shear associated with the large-scale convective columns, rather than by the Reynolds stress due to the parallelization of the small-scale magnetic fields with the small-scale flows.

Our study highlights various impacts that SSD has on the properties of rotating columnar convection. These effects need to be taken into account to properly model rotating magneto-convection particularly in rapidly rotating stars where the SSD significantly alters the dominant force balance. 
Given that fully resolving SSD in global models of solar/stellar convection requires a substantial amount of computational resources, a promising alternative would be to develop subgrid-scale models that encapsulate the essential effects of SSD, as has been attempted in some earlier studies \citep[][]{bekki2017b,hotta2017b,karak2018}.

What are the possible implications for the solar convective conundrum? 
As noted in \S~\ref{sec:intro}, one of the most striking issues is the absence of columnar patterns in the solar surface observations. 
In this work, we have shown that the SSD acts as an effective viscosity on columnar convection, contributing to the suppression of their mode power. In fact, the kinetic energy and enstrophy associated with columnar modes are reduced by approximately 30--50\% due to the SSD effect (see Fig.~\ref{fig:EkinEnst_ky0_vol}). 
However, even in the presence of vigorous SSD, our simulations exhibit highly coherent columnar structures near the solar regime (Figs.~\ref{fig:snap_vz_Co2}--\ref{fig:snap_vz_Co5}). 
Therefore, the SSD effect alone is likely insufficient to fully account for the absence of observational evidence for columnar convection in the Sun. 
Some authors \citep[e.g.,][]{guerrero2013,noraz2025} have suggested that the columnar patterns present in the deep convection zone may be obscured by small-scale granular convection near the surface. The density stratification in our simulations ($N_{\rho} \approx 3.2$) is not strong enough to produce such a concealing effect. Future investigations are required to determine how the results presented would change under conditions of stronger density stratification.


\begin{acknowledgements}
The author appreciates an anonymous referee for the valuable remarks and suggestions.
The author further thanks J.~Wicht, T.~Bhatia, and H.~Hotta for their constructive comments.
Y.~B. acknowledges a support from ERC Synergy Grant WHOLESUN 810218.
All the numerical computations were performed at GWDG and the Max-Planck supercomputer RZG in Garching. 
This work benefited from discussions within the NORDITA workshop "Stellar Convection: Modeling, Theory, and Observations".
\end{acknowledgements}


\begin{thebibliography}{100}
\expandafter\ifx\csname natexlab\endcsname\relax\def\natexlab#1{#1}\fi

\bibitem[{{Aguirre Guzm{\'a}n} {et~al.}(2021){Aguirre Guzm{\'a}n}, {Madonia},
  {Cheng}, {Ostilla-M{\'o}nico}, {Clercx}, \& {Kunnen}}]{guzman2021}
{Aguirre Guzm{\'a}n}, A.~J., {Madonia}, M., {Cheng}, J.~S., {et~al.} 2021, J.
  Fluid. Mech., 928, A16

\bibitem[{{Aubert} {et~al.}(2017){Aubert}, {Gastine}, \&
  {Fournier}}]{aubert2017}
{Aubert}, J., {Gastine}, T., \& {Fournier}, A. 2017, J. Fluid. Mech., 813, 558

\bibitem[{{Augustson} {et~al.}(2015){Augustson}, {Brun}, {Miesch}, \&
  {Toomre}}]{augustson2015}
{Augustson}, K., {Brun}, A.~S., {Miesch}, M., \& {Toomre}, J. 2015, \apj, 809,
  149

\bibitem[{{Augustson} {et~al.}(2019){Augustson}, {Brun}, \&
  {Toomre}}]{augustson2019}
{Augustson}, K.~C., {Brun}, A.~S., \& {Toomre}, J. 2019, \apj, 876, 83

\bibitem[{{Aurnou} {et~al.}(2020){Aurnou}, {Horn}, \& {Julien}}]{aurnou2020}
{Aurnou}, J.~M., {Horn}, S., \& {Julien}, K. 2020, Phy. Rev. Res., 2, 043115

\bibitem[{{Barekat} {et~al.}(2021){Barekat}, {K{\"a}pyl{\"a}},
  {K{\"a}pyl{\"a}}, {Gilson}, \& {Ji}}]{barekat2021}
{Barekat}, A., {K{\"a}pyl{\"a}}, M.~J., {K{\"a}pyl{\"a}}, P.~J., {Gilson},
  E.~P., \& {Ji}, H. 2021, \aap, 655, A79

\bibitem[{{Bekki}(2024)}]{bekki2024a}
{Bekki}, Y. 2024, \aap, 682, A39

\bibitem[{{Bekki} {et~al.}(2022{\natexlab{a}}){Bekki}, {Cameron}, \&
  {Gizon}}]{bekki2022b}
{Bekki}, Y., {Cameron}, R.~H., \& {Gizon}, L. 2022{\natexlab{a}}, \aap, 666,
  A135

\bibitem[{{Bekki} {et~al.}(2022{\natexlab{b}}){Bekki}, {Cameron}, \&
  {Gizon}}]{bekki2022a}
{Bekki}, Y., {Cameron}, R.~H., \& {Gizon}, L. 2022{\natexlab{b}}, \aap, 662,
  A16

\bibitem[{{Bekki} {et~al.}(2017){Bekki}, {Hotta}, \& {Yokoyama}}]{bekki2017b}
{Bekki}, Y., {Hotta}, H., \& {Yokoyama}, T. 2017, \apj, 851, 74

\bibitem[{{Bessolaz} \& {Brun}(2011)}]{bessolaz2011}
{Bessolaz}, N. \& {Brun}, A.~S. 2011, \apj, 728, 115

\bibitem[{{Bhatia} {et~al.}(2022){Bhatia}, {Cameron}, {Solanki}, {Peter},
  {Przybylski}, {Witzke}, \& {Shapiro}}]{bhatia2022}
{Bhatia}, T.~S., {Cameron}, R.~H., {Solanki}, S.~K., {et~al.} 2022, \aap, 663,
  A166

\bibitem[{{Braginskiy}(1970)}]{braginsky1970}
{Braginskiy}, S.~I. 1970, Geomagnetism and Aeronomy, 10, 1

\bibitem[{{Brandenburg}(2016)}]{brandenburg2016}
{Brandenburg}, A. 2016, \apj, 832, 6

\bibitem[{{Brown} {et~al.}(2011){Brown}, {Miesch}, {Browning}, {Brun}, \&
  {Toomre}}]{brown2011}
{Brown}, B.~P., {Miesch}, M.~S., {Browning}, M.~K., {Brun}, A.~S., \& {Toomre},
  J. 2011, \apj, 731, 69

\bibitem[{{Brun} {et~al.}(2004){Brun}, {Miesch}, \& {Toomre}}]{brun2004}
{Brun}, A.~S., {Miesch}, M.~S., \& {Toomre}, J. 2004, \apj, 614, 1073

\bibitem[{{Brun} {et~al.}(2022){Brun}, {Strugarek}, {Noraz}, {Perri}, {Varela},
  {Augustson}, {Charbonneau}, \& {Toomre}}]{brun2022}
{Brun}, A.~S., {Strugarek}, A., {Noraz}, Q., {et~al.} 2022, \apj, 926, 21

\bibitem[{{Brun} \& {Toomre}(2002)}]{brun2002}
{Brun}, A.~S. \& {Toomre}, J. 2002, \apj, 570, 865

\bibitem[{{Bushby} {et~al.}(2018){Bushby}, {K{\"a}pyl{\"a}}, {Masada},
  {Brandenburg}, {Favier}, {Guervilly}, \& {K{\"a}pyl{\"a}}}]{bushby2018}
{Bushby}, P.~J., {K{\"a}pyl{\"a}}, P.~J., {Masada}, Y., {et~al.} 2018, \aap,
  612, A97

\bibitem[{{Busse}(1970)}]{busse1970}
{Busse}, F.~H. 1970, J. Fluid. Mech., 44, 441

\bibitem[{{Busse}(2002)}]{busse2002}
{Busse}, F.~H. 2002, Phys. Fluids, 14, 1301

\bibitem[{{Cattaneo}(1999)}]{cattaneo1999}
{Cattaneo}, F. 1999, \apjl, 515, L39

\bibitem[{{Chandrasekhar}(1961)}]{chandrasekhar1961}
{Chandrasekhar}, S. 1961, {Hydrodynamic and hydromagnetic stability} (Oxford:
  Clarendon)

\bibitem[{{Christensen} \& {Aubert}(2006)}]{christensen2006}
{Christensen}, U.~R. \& {Aubert}, J. 2006, Geophys. J. Int., 166, 97

\bibitem[{{Davidson}(2013)}]{davidson2013}
{Davidson}, P.~A. 2013, Geophys. J. Int., 195, 67

\bibitem[{{Dedner} {et~al.}(2002){Dedner}, {Kemm}, {Kr{\"o}ner}, {Munz},
  {Schnitzer}, \& {Wesenberg}}]{dedner2002}
{Dedner}, A., {Kemm}, F., {Kr{\"o}ner}, D., {et~al.} 2002, J. Comput. Phys.,
  175, 645

\bibitem[{{Evonuk}(2008)}]{evonuk2008}
{Evonuk}, M. 2008, \apj, 673, 1154

\bibitem[{{Fan} \& {Fang}(2014)}]{fan2014}
{Fan}, Y. \& {Fang}, F. 2014, \apj, 789, 35

\bibitem[{{Fan} {et~al.}(1999){Fan}, {Zweibel}, {Linton}, \&
  {Fisher}}]{fan1999}
{Fan}, Y., {Zweibel}, E.~G., {Linton}, M.~G., \& {Fisher}, G.~H. 1999, \apj,
  521, 460

\bibitem[{{Favier} \& {Bushby}(2012)}]{favier2012}
{Favier}, B. \& {Bushby}, P.~J. 2012, J. Fluid. Mech., 690, 262

\bibitem[{{Featherstone} \& {Hindman}(2016)}]{featherstone2016}
{Featherstone}, N.~A. \& {Hindman}, B.~W. 2016, \apjl, 830, L15

\bibitem[{{Featherstone} \& {Miesch}(2015)}]{featherstone2015}
{Featherstone}, N.~A. \& {Miesch}, M.~S. 2015, \apj, 804, 67

\bibitem[{{Foukal} \& {Jokipii}(1975)}]{foukal1975}
{Foukal}, P. \& {Jokipii}, J.~R. 1975, \apjl, 199, L71

\bibitem[{{Gastine} {et~al.}(2013){Gastine}, {Wicht}, \&
  {Aurnou}}]{gastine2013}
{Gastine}, T., {Wicht}, J., \& {Aurnou}, J.~M. 2013, \icarus, 225, 156

\bibitem[{{Gilman} \& {Miller}(1986)}]{gilman1986}
{Gilman}, P.~A. \& {Miller}, J. 1986, \apjs, 61, 585

\bibitem[{{Glatzmaier} {et~al.}(2009){Glatzmaier}, {Evonuk}, \&
  {Rogers}}]{glatzmaier2009}
{Glatzmaier}, G., {Evonuk}, M., \& {Rogers}, T. 2009, Geophys. Astrophys. Fluid
  Dyn., 103, 31

\bibitem[{{Glatzmaier} \& {Gilman}(1981)}]{glatzmaier1981}
{Glatzmaier}, G.~A. \& {Gilman}, P.~A. 1981, \apjs, 45, 381

\bibitem[{{Guerrero} {et~al.}(2013){Guerrero}, {Smolarkiewicz}, {Kosovichev},
  \& {Mansour}}]{guerrero2013}
{Guerrero}, G., {Smolarkiewicz}, P.~K., {Kosovichev}, A.~G., \& {Mansour},
  N.~N. 2013, \apj, 779, 176

\bibitem[{{Hanasoge} {et~al.}(2012){Hanasoge}, {Duvall}, \&
  {Sreenivasan}}]{hanasoge2012}
{Hanasoge}, S.~M., {Duvall}, T.~L., \& {Sreenivasan}, K.~R. 2012, PNAS, 109,
  11928

\bibitem[{{Hanson} {et~al.}(2022){Hanson}, {Hanasoge}, \&
  {Sreenivasan}}]{hanson2022}
{Hanson}, C.~S., {Hanasoge}, S., \& {Sreenivasan}, K.~R. 2022, Nature Astronomy

\bibitem[{{Hathaway} {et~al.}(2015){Hathaway}, {Teil}, {Norton}, \&
  {Kitiashvili}}]{hathaway2015}
{Hathaway}, D.~H., {Teil}, T., {Norton}, A.~A., \& {Kitiashvili}, I. 2015,
  \apj, 811, 105

\bibitem[{{Hindman} \& {Jain}(2022)}]{hindman2022}
{Hindman}, B.~W. \& {Jain}, R. 2022, \apj, 932, 68

\bibitem[{{Hori} {et~al.}(2023){Hori}, {Jones}, {Antu{\~n}ano}, {Fletcher}, \&
  {Tobias}}]{hori2023}
{Hori}, K., {Jones}, C.~A., {Antu{\~n}ano}, A., {Fletcher}, L.~N., \& {Tobias},
  S.~M. 2023, Nature Astronomy, 7, 825

\bibitem[{{Hotta}(2017)}]{hotta2017b}
{Hotta}, H. 2017, \apj, 845, 164

\bibitem[{{Hotta} {et~al.}(2023){Hotta}, {Bekki}, {Gizon}, {Noraz}, \&
  {Rast}}]{hotta2023_review}
{Hotta}, H., {Bekki}, Y., {Gizon}, L., {Noraz}, Q., \& {Rast}, M. 2023, \ssr,
  219, 77

\bibitem[{{Hotta} {et~al.}(2022){Hotta}, {Kusano}, \& {Shimada}}]{hotta2022}
{Hotta}, H., {Kusano}, K., \& {Shimada}, R. 2022, \apj, 933, 199

\bibitem[{{Hotta} {et~al.}(2015{\natexlab{a}}){Hotta}, {Rempel}, \&
  {Yokoyama}}]{hotta2015}
{Hotta}, H., {Rempel}, M., \& {Yokoyama}, T. 2015{\natexlab{a}}, \apj, 803, 42

\bibitem[{{Hotta} {et~al.}(2015{\natexlab{b}}){Hotta}, {Rempel}, \&
  {Yokoyama}}]{hotta2014b}
{Hotta}, H., {Rempel}, M., \& {Yokoyama}, T. 2015{\natexlab{b}}, \apj, 798, 51

\bibitem[{{Hotta} {et~al.}(2016){Hotta}, {Rempel}, \& {Yokoyama}}]{hotta2016}
{Hotta}, H., {Rempel}, M., \& {Yokoyama}, T. 2016, Science, 351, 1427

\bibitem[{{Howard} \& {Labonte}(1980)}]{howard1980}
{Howard}, R. \& {Labonte}, B.~J. 1980, \apjl, 239, L33

\bibitem[{{Jain} \& {Hindman}(2023)}]{jain2023}
{Jain}, R. \& {Hindman}, B.~W. 2023, \apj, 958, 48

\bibitem[{{K{\"a}pyl{\"a}}(2023)}]{kapyla2023}
{K{\"a}pyl{\"a}}, P.~J. 2023, \aap, 669, A98

\bibitem[{{K{\"a}pyl{\"a}}(2024)}]{kapyla2024}
{K{\"a}pyl{\"a}}, P.~J. 2024, \aap, 683, A221

\bibitem[{{K{\"a}pyl{\"a}}(2025)}]{kapyla2025}
{K{\"a}pyl{\"a}}, P.~J. 2025, \aap, 698, L13

\bibitem[{{K{\"a}pyl{\"a}} {et~al.}(2023){K{\"a}pyl{\"a}}, {Browning}, {Brun},
  {Guerrero}, \& {Warnecke}}]{kapyla2023_review}
{K{\"a}pyl{\"a}}, P.~J., {Browning}, M.~K., {Brun}, A.~S., {Guerrero}, G., \&
  {Warnecke}, J. 2023, \ssr, 219, 58

\bibitem[{{K{\"a}pyl{\"a}} {et~al.}(2018){K{\"a}pyl{\"a}}, {K{\"a}pyl{\"a}}, \&
  {Brandenburg}}]{kapyla2018}
{K{\"a}pyl{\"a}}, P.~J., {K{\"a}pyl{\"a}}, M.~J., \& {Brandenburg}, A. 2018,
  Astron. Nachr., 339, 127

\bibitem[{{K{\"a}pyl{\"a}} {et~al.}(2017{\natexlab{a}}){K{\"a}pyl{\"a}},
  {K{\"a}pyl{\"a}}, {Olspert}, {Warnecke}, \& {Brandenburg}}]{kapyla2017b}
{K{\"a}pyl{\"a}}, P.~J., {K{\"a}pyl{\"a}}, M.~J., {Olspert}, N., {Warnecke},
  J., \& {Brandenburg}, A. 2017{\natexlab{a}}, \aap, 599, A4

\bibitem[{{K{\"a}pyl{\"a}} {et~al.}(2009){K{\"a}pyl{\"a}}, {Korpi}, \&
  {Brandenburg}}]{kapyla2009}
{K{\"a}pyl{\"a}}, P.~J., {Korpi}, M.~J., \& {Brandenburg}, A. 2009, \apj, 697,
  1153

\bibitem[{{K{\"a}pyl{\"a}} {et~al.}(2011){K{\"a}pyl{\"a}}, {Mantere}, \&
  {Brandenburg}}]{kapyla2011}
{K{\"a}pyl{\"a}}, P.~J., {Mantere}, M.~J., \& {Brandenburg}, A. 2011, Astron.
  Nachr., 332, 883

\bibitem[{{K{\"a}pyl{\"a}} {et~al.}(2017{\natexlab{b}}){K{\"a}pyl{\"a}},
  {Rheinhardt}, {Brand enburg}, {Arlt}, {K{\"a}pyl{\"a}}, {Lagg}, {Olspert}, \&
  {Warnecke}}]{kapyla2017}
{K{\"a}pyl{\"a}}, P.~J., {Rheinhardt}, M., {Brand enburg}, A., {et~al.}
  2017{\natexlab{b}}, \apjl, 845, L23

\bibitem[{{K{\"a}pyl{\"a}} {et~al.}(2019){K{\"a}pyl{\"a}}, {Viviani},
  {K{\"a}pyl{\"a}}, {Brandenburg}, \& {Spada}}]{kapyla2019}
{K{\"a}pyl{\"a}}, P.~J., {Viviani}, M., {K{\"a}pyl{\"a}}, M.~J., {Brandenburg},
  A., \& {Spada}, F. 2019, Geophys. Astrophys. Fluid Dyn., 113, 149

\bibitem[{{Karak} {et~al.}(2015){Karak}, {K{\"a}pyl{\"a}}, {K{\"a}pyl{\"a}},
  {Brandenburg}, {Olspert}, \& {Pelt}}]{karak2015}
{Karak}, B.~B., {K{\"a}pyl{\"a}}, P.~J., {K{\"a}pyl{\"a}}, M.~J., {et~al.}
  2015, \aap, 576, A26

\bibitem[{{Karak} {et~al.}(2018){Karak}, {Miesch}, \& {Bekki}}]{karak2018}
{Karak}, B.~B., {Miesch}, M., \& {Bekki}, Y. 2018, Phys. Fluids, 30, 046602

\bibitem[{{Kitchatinov}(2013)}]{kitchatinov2013}
{Kitchatinov}, L.~L. 2013, in IAU Symposium, Vol. 294, Solar and Astrophysical
  Dynamos and Magnetic Activity, ed. A.~G. {Kosovichev}, E.~{de Gouveia Dal
  Pino}, \& Y.~{Yan}, 399--410

\bibitem[{{Mabuchi} {et~al.}(2015){Mabuchi}, {Masada}, \&
  {Kageyama}}]{mabuchi2015}
{Mabuchi}, J., {Masada}, Y., \& {Kageyama}, A. 2015, \apj, 806, 10

\bibitem[{{Masada} \& {Sano}(2014)}]{masada2014b}
{Masada}, Y. \& {Sano}, T. 2014, \pasj, 66, S2

\bibitem[{{Matilsky} {et~al.}(2022){Matilsky}, {Hindman}, {Featherstone},
  {Blume}, \& {Toomre}}]{matilsky2022}
{Matilsky}, L.~I., {Hindman}, B.~W., {Featherstone}, N.~A., {Blume}, C.~C., \&
  {Toomre}, J. 2022, \apjl, 940, L50

\bibitem[{{Matilsky} {et~al.}(2019){Matilsky}, {Hindman}, \&
  {Toomre}}]{matilsky2019}
{Matilsky}, L.~I., {Hindman}, B.~W., \& {Toomre}, J. 2019, \apj, 871, 217

\bibitem[{{Matilsky} {et~al.}(2020){Matilsky}, {Hindman}, \&
  {Toomre}}]{matilsky2020}
{Matilsky}, L.~I., {Hindman}, B.~W., \& {Toomre}, J. 2020, \apj, 898, 111

\bibitem[{{Miesch}(2005)}]{miesch2005}
{Miesch}, M.~S. 2005, Living Reviews in Solar Physics, 2, 1

\bibitem[{{Miesch} {et~al.}(2008){Miesch}, {Brun}, {DeRosa}, \&
  {Toomre}}]{miesch2008}
{Miesch}, M.~S., {Brun}, A.~S., {DeRosa}, M.~L., \& {Toomre}, J. 2008, \apj,
  673, 557

\bibitem[{{Miesch} {et~al.}(2000){Miesch}, {Elliott}, {Toomre}, {Clune},
  {Glatzmaier}, \& {Gilman}}]{miesch2000}
{Miesch}, M.~S., {Elliott}, J.~R., {Toomre}, J., {et~al.} 2000, \apj, 532, 593

\bibitem[{{Mori} \& {Hotta}(2023)}]{mori2023a}
{Mori}, K. \& {Hotta}, H. 2023, \mnras, 519, 3091

\bibitem[{{Nelson} {et~al.}(2013){Nelson}, {Brown}, {Brun}, {Miesch}, \&
  {Toomre}}]{nelson2013}
{Nelson}, N.~J., {Brown}, B.~P., {Brun}, A.~S., {Miesch}, M.~S., \& {Toomre},
  J. 2013, \apj, 762, 73

\bibitem[{{Noraz} {et~al.}(2025){Noraz}, {Brun}, \& {Strugarek}}]{noraz2025}
{Noraz}, Q., {Brun}, A.~S., \& {Strugarek}, A. 2025, \apj, 981, 206

\bibitem[{{O'Mara} {et~al.}(2016){O'Mara}, {Miesch}, {Featherstone}, \&
  {Augustson}}]{omara2016}
{O'Mara}, B., {Miesch}, M.~S., {Featherstone}, N.~A., \& {Augustson}, K.~C.
  2016, Advances in Space Research, 58, 1475

\bibitem[{{Ong} \& {Roundy}(2020)}]{ong2020}
{Ong}, H. \& {Roundy}, P.~E. 2020, J. Atmos. Sci., 77, 3721

\bibitem[{{Ossendrijver}(2003)}]{ossendrijver2003}
{Ossendrijver}, M. 2003, \aapr, 11, 287

\bibitem[{{Rempel}(2014)}]{rempel2014}
{Rempel}, M. 2014, \apj, 789, 132

\bibitem[{{Rempel}(2018)}]{rempel2018}
{Rempel}, M. 2018, \apj, 859, 161

\bibitem[{{Rempel} {et~al.}(2023){Rempel}, {Bhatia}, {Bellot Rubio}, \&
  {Korpi-Lagg}}]{rempel2023_review}
{Rempel}, M., {Bhatia}, T., {Bellot Rubio}, L., \& {Korpi-Lagg}, M.~J. 2023,
  \ssr, 219, 36

\bibitem[{Roberts(1978)}]{roberts1978}
Roberts, P. 1978, Rotating Fluids in Geophysics (ed. PH Roberts \& AM Soward),
  421

\bibitem[{{Rogachevskii} \& {Kleeorin}(2003)}]{rogachevskii2003}
{Rogachevskii}, I. \& {Kleeorin}, N. 2003, \pre, 68, 036301

\bibitem[{{Schwaiger} {et~al.}(2021){Schwaiger}, {Gastine}, \&
  {Aubert}}]{schwaiger2021}
{Schwaiger}, T., {Gastine}, T., \& {Aubert}, J. 2021, Geophys. J. Int., 224,
  1890

\bibitem[{{Soderlund} {et~al.}(2025){Soderlund}, {Wulff}, {K{\"a}pyl{\"a}}, \&
  {Aurnou}}]{soderlund2025}
{Soderlund}, K.~M., {Wulff}, P., {K{\"a}pyl{\"a}}, P.~J., \& {Aurnou}, J.~M.
  2025, \mnras, 541, 1816

\bibitem[{{Strugarek} {et~al.}(2017){Strugarek}, {Beaudoin}, {Charbonneau},
  {Brun}, \& {do Nascimento}}]{strugarek2017}
{Strugarek}, A., {Beaudoin}, P., {Charbonneau}, P., {Brun}, A.~S., \& {do
  Nascimento}, J.~D. 2017, Science, 357, 185

\bibitem[{{Takehiro}(2008)}]{takehiro2008}
{Takehiro}, S.-I. 2008, J. Fluid. Mech., 614, 67

\bibitem[{Taylor(1963)}]{taylor1963}
Taylor, J.~B. 1963, Proceedings of the Royal Society of London. Series A.
  Mathematical and Physical Sciences, 274, 274

\bibitem[{{Thompson} {et~al.}(1996){Thompson}, {Toomre}, {Anderson}, {Antia},
  {Berthomieu}, {Burtonclay}, {Chitre}, {Christensen-Dalsgaard}, {Corbard}, {De
  Rosa}, {Genovese}, {Gough}, {Haber}, {Harvey}, {Hill}, {Howe}, {Korzennik},
  {Kosovichev}, {Leibacher}, {Pijpers}, {Provost}, {Rhodes}, {Schou}, {Sekii},
  {Stark}, \& {Wilson}}]{thompson1996}
{Thompson}, M.~J., {Toomre}, J., {Anderson}, E.~R., {et~al.} 1996, Science,
  272, 1300

\bibitem[{{Vasavada} \& {Showman}(2005)}]{vasavada2005}
{Vasavada}, A.~R. \& {Showman}, A.~P. 2005, Reports on Progress in Physics, 68,
  1935

\bibitem[{Vasil {et~al.}(2021)Vasil, Julien, \& Featherstone}]{vasil2021}
Vasil, G.~M., Julien, K., \& Featherstone, N.~A. 2021, PNAS, 118, e2022518118

\bibitem[{{Verhoeven} \& {Stellmach}(2014)}]{verhoeven2014}
{Verhoeven}, J. \& {Stellmach}, S. 2014, \icarus, 237, 143

\bibitem[{{Viviani} {et~al.}(2018){Viviani}, {Warnecke}, {K{\"a}pyl{\"a}},
  {K{\"a}pyl{\"a}}, {Olspert}, {Cole-Kodikara}, {Lehtinen}, \&
  {Brandenburg}}]{viviani2018}
{Viviani}, M., {Warnecke}, J., {K{\"a}pyl{\"a}}, M.~J., {et~al.} 2018, \aap,
  616, A160

\bibitem[{{V{\"o}gler} {et~al.}(2005){V{\"o}gler}, {Shelyag}, {Sch{\"u}ssler},
  {Cattaneo}, {Emonet}, \& {Linde}}]{voegler2005}
{V{\"o}gler}, A., {Shelyag}, S., {Sch{\"u}ssler}, M., {et~al.} 2005, \aap, 429,
  335

\bibitem[{{Warnecke} {et~al.}(2023){Warnecke}, {Korpi-Lagg}, {Gent}, \&
  {Rheinhardt}}]{warnecke2023}
{Warnecke}, J., {Korpi-Lagg}, M.~J., {Gent}, F.~A., \& {Rheinhardt}, M. 2023,
  Nature Astronomy, 7, 662

\bibitem[{{Warnecke} {et~al.}(2025){Warnecke}, {Korpi-Lagg}, {Rheinhard},
  {Viviani}, \& {Prabhu}}]{warnecke2025}
{Warnecke}, J., {Korpi-Lagg}, M.~J., {Rheinhard}, M., {Viviani}, M., \&
  {Prabhu}, A. 2025, \aap, 696, A93

\bibitem[{{Wicht} \& {Christensen}(2010)}]{wicht2010}
{Wicht}, J. \& {Christensen}, U.~R. 2010, Geophys. J. Int., 181, 1367

\bibitem[{{Yadav} {et~al.}(2016){Yadav}, {Gastine}, {Christensen}, {Wolk}, \&
  {Poppenhaeger}}]{yadav2016}
{Yadav}, R.~K., {Gastine}, T., {Christensen}, U.~R., {Wolk}, S.~J., \&
  {Poppenhaeger}, K. 2016, PNAS, 113, 12065

\bibitem[{Yan {et~al.}(2021)Yan, Tobias, \& Calkins}]{yan2021}
Yan, M., Tobias, S., \& Calkins, M. 2021, J. Fluid Mech., 915, A15

\bibitem[{{Yousef} {et~al.}(2008){Yousef}, {Heinemann}, {Schekochihin},
  {Kleeorin}, {Rogachevskii}, {Iskakov}, {Cowley}, \&
  {McWilliams}}]{yousef2008}
{Yousef}, T.~A., {Heinemann}, T., {Schekochihin}, A.~A., {et~al.} 2008, \prl,
  100, 184501

\end{thebibliography}


\clearpage

\begin{appendix} 


\section{Numerical diffusion scheme} \label{appendix:numerical_diffusion}

In this study, we do not use any explicit diffusivities but instead employ a slope-limited artificial diffusion scheme \citep{rempel2014}.
This Appendix outlines how this artificial diffusion is implemented and how the effective diffusivities are estimated in our simulations.

The numerical diffusive terms are added in Eqs.~(\ref{eq:conti})--(\ref{eq:ent}) as
\begin{eqnarray}
    && \frac{\partial \rho_1}{\partial t} = [...]-\nabla\cdot \bm{f}^{\rm diff}(\rho_1), \\
    && \rho_0 \frac{\partial \bm{v}}{\partial t}=[...]-\nabla\cdot [\rho_0 \bm{f}^{\rm diff}(\bm{v})], \\
    && \frac{\partial \bm{B}}{\partial t}=[...]-\nabla \cdot \bm{f}^{\rm diff}(\bm{B}), \\
    && \rho_0 T_0 \frac{\partial s_1}{\partial t} = [...] - \nabla\cdot[\rho_0 T_0 \bm{f}^{\rm diff}(s_1)] + Q^{\rm diff}_{\rm visc} + Q^{\rm diff}_{\rm joul}, 
\end{eqnarray}
where $\bm{f}^{\rm diff}(q)$ represents the numerical diffusive flux of a variable $q$ which is non-linearly extrapolated at the grid interfaces \citep[see \S~2.1 in][for its detailed form]{rempel2014}.
For a parameter controlling the strength of diffusion \citep[expressed in Eq.10 in][]{rempel2014}, we set $h=1.5$.
In the entropy equation, we add the heating terms arising from the kinetic and magnetic energy dissipation, $Q^{\rm diff}_{\rm visc}$ and $Q^{\rm diff}_{\rm joul}$. They are estimated as
\begin{eqnarray}
    && Q^{\rm diff}_{\rm visc} = -\rho_0 \sum_{i,k} \frac{\partial v_i}{\partial x_k} f^{\rm diff}_{k}(v_i), \\
    && Q^{\rm diff}_{\rm joul} = -\frac{1}{4\pi} \sum_{i,k} \frac{\partial B_i}{\partial x_k} f^{\rm diff}_{k}(B_i).
\end{eqnarray}

To infer the effective viscous and magnetic diffusivities, $\nu_{\rm eff}$ and $\eta_{\rm eff}$ in our simulations, we compare the above numerical energy dissipation rates with those inferred from explicit diffusivities as \citep{hotta2017b,rempel2018}
\begin{eqnarray}
    &&  \nu_{\rm eff} = \frac{ \langle Q^{\rm diff}_{\rm visc} \rangle}{\rho_0 \Bigl< \sum_{i,k} \frac{\partial v_{i}}{\partial x_{k}} \left[ \frac{\partial v_{i}}{\partial x_{k}} + \frac{\partial v_{k}}{\partial x_{i}} - \frac{2}{3} \delta_{ik}(\nabla\cdot\bm{v}) \right]  \Bigr>}, \\
    && \eta_{\rm eff}= \frac{\langle Q^{\rm diff}_{\rm joul} \rangle}{\Bigl< |\nabla\times\bm{B}|^2 \Bigr>/4\pi }.
\end{eqnarray}
Using the volume-averaged values of these diffusivities, we can estimate the effective Reynolds and magnetic Reynolds numbers, Re$_{*}$ and Rm$_{*}$, as
\begin{eqnarray}
    && \mathrm{Re}_{*}=\frac{v_{*} H_r}{\overline{\nu}_{\rm eff}}, \ \ \ \mathrm{and} \ \ \ \mathrm{Rm}_{*}=\frac{v_{*} H_r}{\overline{\eta}_{\rm eff}}. \label{eq:ReRm}
\end{eqnarray}
The computed values are reported in Table~\ref{table:1}.

\section{Scaling laws in CIA and MAC balances} \label{appendix:scaling}

In this Appendix, we derive ideal scaling relations in rapidly rotating regime for convective velocity, length scale, and entropy perturbation in terms of the flux Coriolis number Co$_*$.
Two different force balances are considered: the Coriolis-Inertial-Archimedean (CIA) balance and the Magneto-Archimedean-Coriolis (MAC) balance.
We analyze the balance in the vorticity equation which can be obtained by taking a curl of Eq.~(\ref{eq:motion}),
\begin{eqnarray}
&& \frac{\partial \bm{\zeta}}{\partial t}=-\bm{v}\cdot \nabla\bm{\zeta}+(\bm{\zeta}+2\bm{\Omega}_{0})\cdot \nabla \bm{v}-(\bm{\zeta}+2\bm{\Omega}_{0})(\nabla\cdot\bm{v}) \nonumber \\ 
&& \ \ \ \ \ \ \ \ 
+\nabla\times \left(\frac{s_{1}}{c_{p}} \right)g \bm{e}_{z} 
+\nabla\times \left[ \frac{1}{4\pi\rho_0} (\nabla\times\bm{B})\times\bm{B} \right]. \label{eq:vort}
\end{eqnarray}

\subsection{CIA balance}

We consider the balance among the advective (inertial), Coriolis, and buoyancy (Archimedean) terms in the $y$-component of Eq.~(\ref{eq:vort}).
Noting that the $y$-component of the Coriolis force is $-2\Omega_0 (\nabla\cdot \bm{v}) = -2\Omega_0 H_\rho^{-1}v_z$, we have 
\begin{eqnarray}
    && \frac{v_{\rm cia}^2}{\ell_{\rm cia}^2} \approx \frac{2\Omega_0 v_{\rm cia}}{H_r} \approx  \frac{g s_{\rm cia}}{c_p \ell_{\rm cia}}, \label{cia-balance}
\end{eqnarray}
where $v_{\rm cia}$, $\ell_{\rm cia}$, and $s_{\rm cia}$ represent the typical convective velocity, horizontal length scale, and entropy fluctuation in the CIA balance.
It is convenient to rewrite the buoyancy term as
\begin{eqnarray}
    &&     \frac{g s_{\rm cia}}{c_p \ell_{\rm cia}} \approx \frac{v_{*}^3}{H_r v_{\rm cia}\ell_{\rm cia}}. \label{eq:cia-buo}
\end{eqnarray}
Here, we assume that the enthalpy flux becomes equal to the injected energy flux in a rapidly rotating limit, $F_{\rm enth} \approx \rho_r T_r v_{\rm cia} s_{\rm cia} \approx F_*$, and use the relation $gH_r = p_r/\rho_r \approx c_p T_r$.
Now, we first consider the CI balance to yield
\begin{eqnarray}
    && \left(\frac{\ell_{\rm cia}}{H_r}\right)^2 \approx  \mathrm{Co}_*^{-1} \left( \frac{v_{\rm cia}}{v_*} \right). \label{eq:ci}
\end{eqnarray}
A similar consideration of the CA balance gives us
\begin{eqnarray}
    && \frac{\ell_{\rm cia}}{H_r} \approx \mathrm{Co}_*^{-1} \left( \frac{v_{\rm cia}}{v_*}\right)^{-2}. \label{eq:ca}
\end{eqnarray}
Combining Eqs.~(\ref{eq:ci}) and (\ref{eq:ca}) yields the scaling relations for velocity $v_{\rm cia}$ and for length scale $\ell_{\rm cia}$
\begin{eqnarray}
    && \frac{v_{\rm cia}}{v_*} \approx \mathrm{Co}_{*}^{-1/5}, \label{eq:CIA-vel} \\
   && \frac{\ell_{\rm cia}}{H_r} \approx  \mathrm{Co}_{*}^{-3/5}. \label{eq:CIA-length}
\end{eqnarray}
Finally, substituting Eqs.~(\ref{eq:CIA-vel} and \ref{eq:CIA-length}) into Eq.~(\ref{eq:cia-buo}) leads to
\begin{eqnarray}
    && \frac{s_{\rm cia}}{c_p} \approx \mathrm{M}_*^2 \mathrm{Co}_{*}^{1/5}.
\end{eqnarray}
These results are consistent with those reported in previous studies \citep[][]{aurnou2020,vasil2021,kapyla2024} where the dependences on the bulk Coriolis number Co$=2\Omega_0 H_r/v_{\rm cia}$ were derived instead, e.g., $\ell_{\rm cia} \propto \mathrm{Co}^{-1/2}$. 

\subsection{MAC balance}

The force balance between Coriolis, buoyancy, and Lorentz forces is called MAC balance \citep[e.g.,][]{davidson2013,yadav2016,augustson2019,schwaiger2021}.
Amplitudes of these forces in the vorticity Eq.~(\ref{eq:vort}) are expressed as
\begin{eqnarray}
    && \frac{2\Omega_0 v_{\rm mac}}{H_r} \approx  \frac{g s_{\rm mac}}{c_p \ell_{\rm mac}} \approx \frac{B_{\rm mac}^2}{4\pi\rho_r \ell_{\rm mac}^2}, \label{eq:mac-balance}
\end{eqnarray}
where $v_{\rm mac}$, $\ell_{\rm mac}$, $s_{\rm mac}$, and $B_{\rm mac}$ represent the typical convective velocity, length scale, entropy fluctuation, and magnetic field strength in MAC balance.
Similarly to the CIA balance, the CA balance in Eq.~(\ref{eq:mac-balance}) can be written as
\begin{eqnarray}
    && \frac{\ell_{\rm mac}}{H_r} \approx \mathrm{Co}_*^{-1} \left( \frac{v_{\rm mac}}{v_*}\right)^{-2}. \label{eq:ca2}
\end{eqnarray}
Now, we need to estimate the typical magnetic field strength $B_{\rm mac}$.
Since the SSD is expected to operate at sufficiently small scales where the turbulent eddy motions are unaffected by rotation, we assume here that the magnetic energy is independent of Co$_*$ \citep[][]{davidson2013}.
In such a case, $B^2_{\rm mac}/4\pi \rho_r$ is solely generated by small-scale convection driven by the injected energy flux $F_*$, and thus is conventionally expressed as
\begin{eqnarray}
    && \frac{B_{\rm mac}}{4\pi \rho_r} \approx v_{*}^2. \label{eq:Bmac}
\end{eqnarray}
Considering the MC balance leads to
\begin{eqnarray}
    && \left(\frac{\ell_{\rm mac}}{H_r}\right)^2 \approx  \mathrm{Co}_*^{-1} \left( \frac{v_{\rm mac}}{v_*} \right)^{-1}. \label{eq:mc}
\end{eqnarray}
From Eqs.~(\ref{eq:ca2})--(\ref{eq:mc}), we have the following scaling laws
\begin{eqnarray}
    && \frac{v_{\rm mac}}{v_*} \approx \mathrm{Co}_*^{-1/3}, \\
    && \frac{\ell_{\rm mac}}{H_r} \approx \mathrm{Co}_*^{-1/3}, \\
    && \frac{s_{\rm mac}}{c_p} \approx \mathrm{M}_*^2 \mathrm{Co}_*^{1/3}.
\end{eqnarray}
It is worth noting that, in terms of the bulk Coriolis number Co$=2\Omega_0 H_r/v_{\rm mac}$, the typical convective length scale follows $\ell_{\rm mac} \propto \mathrm{Co}^{-1/4}$ \citep[e.g.,][]{schwaiger2021}.

\subsection{Physical interpretation}

A relation that is common to both CIA and MAC balances (i.e., the CA balance) is given by
\begin{eqnarray}
 &&   v^2 \ell \propto \mathrm{Co}_*^{-1} \ \ \ \mathrm{with} \ \ \ v \propto s^{-1}.
\end{eqnarray}
Therefore, as rotational influence increases, $v^2 \ell$ must be decreased by suppressing both convective velocity $v$ and length scale $\ell$.
The suppression of $v$ with increasing Co$_*$ can be physically explained as follows: 
Strong rotation reduces the efficiency of convective heat transport (suppresses vertical mixing of entropy), leading to a mean background stratification being more superadiabatic (see Fig.~\ref{fig:delta}). 
In other words, under rapid rotation, the energy that would otherwise be used to increase the kinetic energy of convection by buoyancy is instead retained as internal potential energy in the background stratification. 
We note that this superadiabatic stratification does not result in a stronger convective driving because the buoyancy is effectively balanced by Coriolis force (i.e., CA balance) and can hardly be used to accelerate convection.
When the SSD exists, the kinetic energy is further reduced due to a conversion into magnetic energy.
Therefore, for a fixed Co$_*$, the length scale $\ell$ must increase in the MAC balance in order to compensate for the reduction in $v$ and keep $v^2 \ell$ constant.



\section{Torsional oscillations} \label{appendix:TO}

\begin{figure}
\begin{center}
\includegraphics[width=0.95\linewidth]{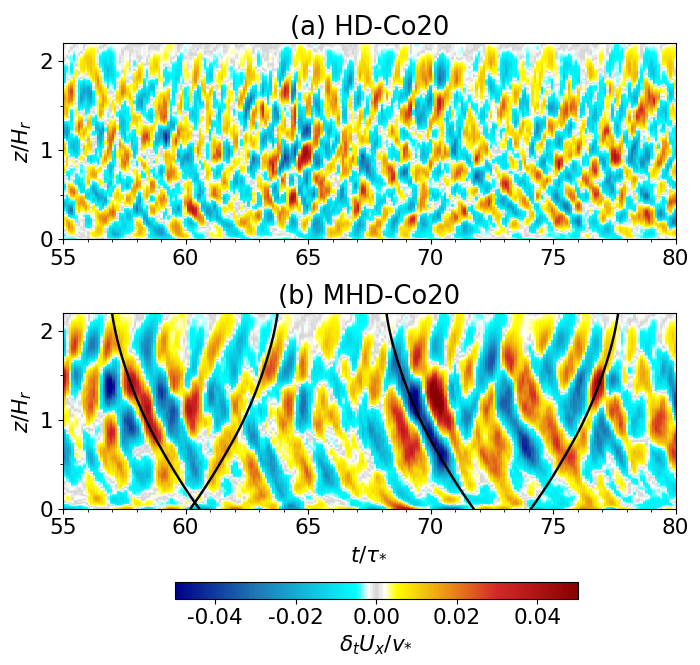}
\caption{
Torsional oscillation, i.e., temporal variation of the longitudinal ($x$) component of the mean flow $\delta_t U_{x}(z,t)$ where the temporal average is subtracted, from  (a) HD-Co20 and (b) MHD-Co20 cases.
Black solid curves showcase paths of waves propagating upward and downward with the Alfvén velocity of the rms vertical field $v_{\mathrm{A},z}=\sqrt{\langle B_z^2 \rangle /4\pi\rho_0}$.
}
\label{fig:TO}
\end{center}
\end{figure}

In the regime where the MAC balance is established, torsional Alfvén waves are known to propagate in a vertical direction as disturbances to the Taylor state \citep[][]{taylor1963,braginsky1970}. 
These waves can be observed as temporal variations of the mean flows known as \textit{torsional oscillations} \citep[e.g.,][]{wicht2010}.
We note that, in solar physics, torsional oscillations conventionally refer to a temporal variation of the Sun's differential rotation caused by the $11$-year cyclic behavior of the solar large-scale dynamo \citep[e.g.,][]{howard1980}.
In our simulations, however, the torsional oscillations occur on much shorter time scales comparable to the convective turnover time scale $\tau_{*}$.

Taking a temporal derivative of Eq.~(\ref{eq:dUxdt}), we have
\begin{eqnarray}
    && \rho_{0} \frac{\partial^{2} U_{x}}{\partial t^{2}} = -\frac{\partial}{\partial z} \left[ \frac{\partial}{\partial t} \left( \rho_0 \langle v'_x v'_z \rangle -\frac{ \langle B_{x}B_{z} \rangle}{4\pi} \right) \right]. \label{eq:d2Uxdt}
\end{eqnarray}
To estimate the temporal variation of the Reynolds and Maxwell stresses, we first consider the terms associated with the mean flow variation in Eqs.~(\ref{eq:motion}) and (\ref{eq:induction}) to obtain
\begin{eqnarray}
    && \frac{\partial}{\partial t} \left( v'_{x}v'_{z}\right) \approx -v_{z}^{2} \frac{\partial U_{x}}{\partial z} -U_{x} \frac{\partial}{\partial x}\left( v'_{x}v'_{z} \right), \label{eq:dRxzdt} \\
    && \frac{\partial}{\partial t} \left( B_{x}B_{z}\right) \approx B_{z}^{2} \frac{\partial U_{x}}{\partial z} -U_{x} \frac{\partial}{\partial x}\left( B_{x}B_{z} \right). \label{eq:dMxzdt}
\end{eqnarray}
Taking a horizontal average of Eqs.~(\ref{eq:dRxzdt} and \ref{eq:dMxzdt}) and substituting them into Eq.~(\ref{eq:d2Uxdt}), we have
\begin{eqnarray}
    && \rho_0 \frac{\partial^{2} U_{x}}{\partial t^{2}} \approx  \frac{\partial}{\partial z} \left[\left( \rho_0 \langle v_z^2 \rangle + \frac{\langle B_{z}^{2} \rangle}{4\pi}  \right) \frac{\partial U_{x}}{\partial z} \right].
\end{eqnarray}
If we assume that $\rho_0 \langle v_z^2\rangle$ and $\langle B_{z}^{2} \rangle$ vary more slowly in $z$ than $\partial U_{x}/\partial z$, the above equation can be reduced to a standard one-dimensional wave equation
\begin{eqnarray}
    && \frac{\partial^{2} U_{x}}{\partial t^{2}} \approx (\langle v_{z}^{2}\rangle + v_{\mathrm{A},z}^{2} )\frac{\partial^{2} U_{x}}{\partial z^{2}}, \ \ \ \mathrm{with} \ \ v_{\mathrm{A},z}^{2} = \frac{\langle B_{z}^{2}\rangle}{4\pi\rho_{0}}, \label{eq:TO}
\end{eqnarray}
where $v_{\mathrm{A},z}$ is the Alfvén velocity associated with the rms amplitude of the vertical magnetic field.
If the MAC balance holds, the above equation (\ref{eq:TO}) is reduced to the equation of torsional Alfvén waves whose propagation leads to a torsional oscillation restored by the Lorentz force of the fields perpendicular to the rotational axis \citep{braginsky1970}.

Figure~\ref{fig:TO} shows time-height plots of the torsional oscillation defined as
\begin{eqnarray}
    && \delta_t U_x (z,t) = U_x(z,t) - \frac{1}{T}\int U_x(z,t) dt,
\end{eqnarray}
for the cases HD-Co20 and MHD-Co20.
Here, the temporal average is taken over a period $T=25 \ \tau_*$.
In the HD-Co20 case, the temporally fluctuating component of the mean flow is weak and the torsional oscillation patterns can hardly be identified with significant signal-to-noise ratio.
In the MHD-Co20 case, on the other hand, we can unambiguously identify the torsional oscillation patterns in Fig.~\ref{fig:TO}b where the disturbances in the mean flow propagate vertically upward or downward with the Alfvén velocity $v_{\rm A,z}$.
The typical amplitude of the torsional oscillation is $\delta_t U_x /v_* \approx 0.05$, which is approximately a few percent of the rms velocity fluctuation.
It remains to be investigated how important these torsional oscillations are to the flow dynamics \citep[e.g.,][]{hori2023}.

\section{Physical origin of the Reynolds stress $\langle v_{x}' v_{z}'\rangle$} \label{appendix:origin_tilt}

\begin{figure}
\begin{center}
\includegraphics[width=\linewidth]{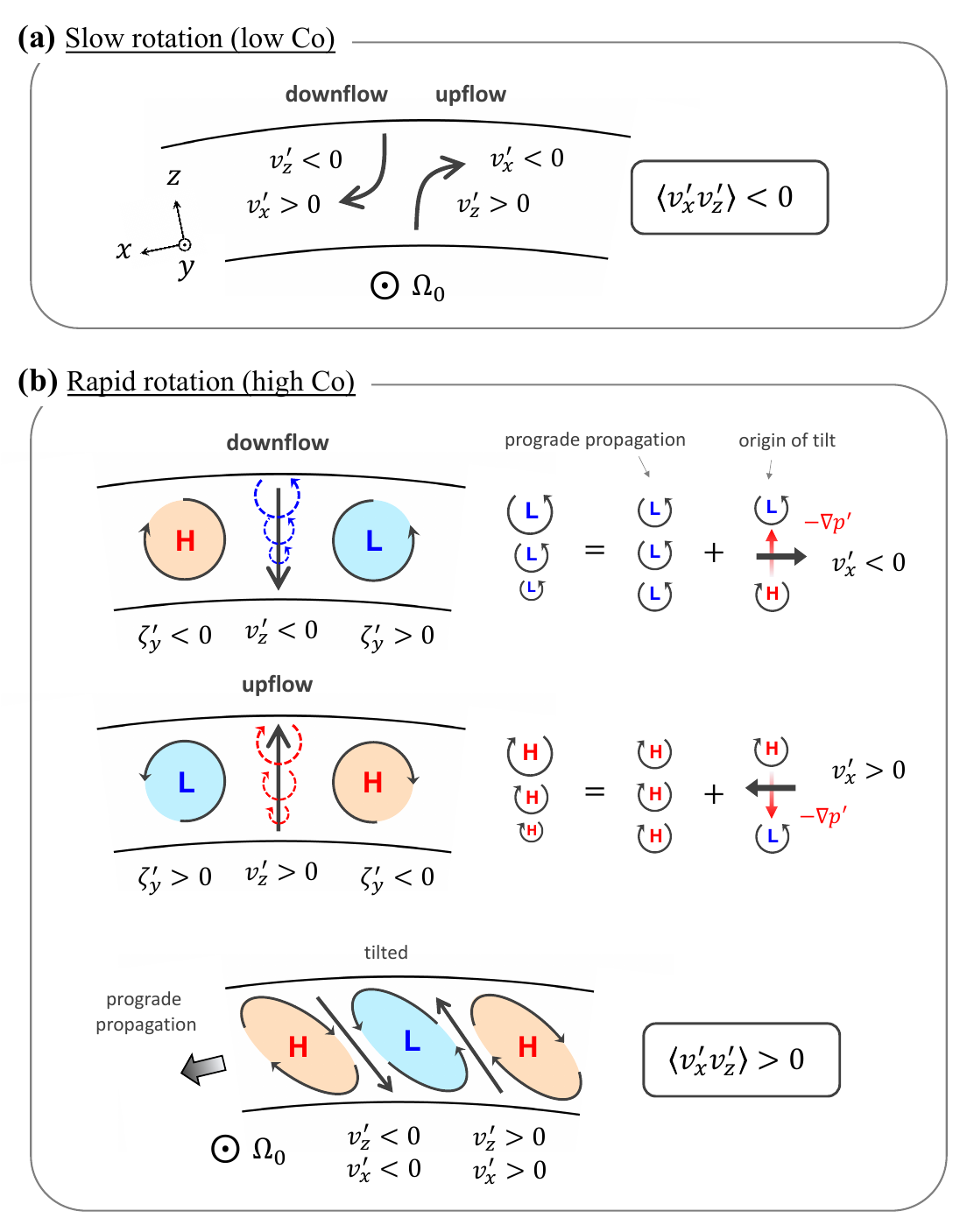}
\caption{
Schematic illustration of the generation mechanisms of the Reynolds stress $\langle v_{x}' v_{z}'\rangle$ under (a) weak and (b) strong rotational influences.
Here, the $x$ and $z$ axes correspond to the longitudinal and radial directions.
All equatorial cuts are viewed from the north pole.
In panel b, orange and cyan circles denote the clockwise ($\zeta'_{y}<0$) and counter-clockwise ($\zeta'_{y}>0$) vortices with positive and negative pressure anomalies.
}
\label{fig:app_origin_tilt}
\end{center}
\end{figure}

In this Appendix, we provide a physical explanation of the generation mechanism of the Reynolds stress $\langle v_{x}' v_{z}' \rangle$ in slowly and rapidly rotating regimes.

We first consider a weak rotation regime (low-Co$_*$ regime).
The Coriolis term in the equation of motion can be expressed as
\begin{eqnarray}
    && \frac{\partial v_{x}'}{\partial t} = [...] -2\Omega_{0} v_{z}', \label{eq:app1} \\
    && \frac{\partial v_{z}'}{\partial t} = [...]+2\Omega_{0} v_{x}'. \label{eq:app2}
\end{eqnarray}
Therefore, the sign of the Reynolds stress depends on whether the Coriolis force acts on vertical velocity $v_{z}'$ or on longitudinal velocity $v_{x}'$.
Since vertical convective motions dominate over horizontal motions at low-Co$_*$ regime, negative Reynolds stress $\langle v_{x}' v_{z}' \rangle <0$ is selectively generated by Coriolis force acting on $v_{z}'$.
This is schematically illustrated in Fig.~\ref{fig:app_origin_tilt}a.

Under a strong rotational influence (high-Co$_*$ regime), on the other hand, the above explanation cannot be applied because the geostrophic balance between the Coriolis force and the pressure gradient force is established.
Since such a flow becomes invariant along the rotational ($y$) axis, it is convenient to consider the $y$-component of the vorticity $\zeta_{y}'=\partial v_{x}'/\partial z -\partial v_{z}'/\partial x$.
Taking a curl of Eqs.~(\ref{eq:app1} and \ref{eq:app2}), we have
\begin{eqnarray}
 && \frac{\partial \zeta_{y}'}{\partial t} = [...]-\frac{2\Omega_{0}}{H_{\rho}} v_{z}' .   \label{eq:app3}
\end{eqnarray}
This tells us that, due to the background density stratification, vertical motions cause an expansion or contraction of $y$-vortices.
The generation of positive (negative) $y$-vorticity $\zeta_{y}'>0 \ (<0)$ in downflow (upflow) regions results in prograde propagation of the overall $y$-vortex columns, i.e., thermal Rossby waves.
Now, a key to understanding the origin of the Reynolds stress is that the generation rate of this $y$-vorticity perturbation is stronger near the surface (where the density scale height $H_{\rho}$ is smaller) as expressed in Eq.~(\ref{eq:app3}).
This causes vertically upward (downward) pressure gradient forces in downflow (upflow) regions because, in a geostrophic fluid, positive (negative) vortices $\zeta_{y}'>0 \ (<0)$ are accompanied by negative (positive) pressure anomalies $p'<0 \ (>0)$.
In order to balance these additional vertical pressure gradient forces by the Coriolis force, retrograde (prograde) velocity perturbations $v_{x}' <0 \ (>0)$ are required in downflow (upflow) regions.
Therefore, upflows and downflows are deflected to prograde and retrograde directions, respectively, leading to a net positive Reynolds stress $\langle v_{x}' v_{z}' \rangle >0$.
This positive Reynolds stress can be seen as a tilted pattern of the vortex columns, as illustrated in Fig.~\ref{fig:app_origin_tilt}b.

We note that the above explanation is similar to but different from the conventional explanation that the vortex columns are tilted in a prograde direction because the prograde phase speed of the thermal Rossby waves is faster near the surface than near the base \citep[e.g.,][]{busse2002}.
Although this explanation is widely used within the community \citep[e.g.,][]{vasavada2005,glatzmaier2009}, it has sometimes been criticized as unsatisfactory because the tilted pattern can be obtained as a linear eigenfunction of a global (normal) mode which propagates at a fixed phase speed everywhere \citep[][]{takehiro2008}.
Here, we provide an updated explanation for the origin of the Reynolds stress which does not rely on any nonlinear processes and thus is more satisfactory.
Finally, it should be noted that our explanation can also be applied to incompressible thermal Rossby waves caused by the spherical curvature effect \citep[e.g.,][]{busse2002} by simply replacing the compressional $\beta$-effect, $-2\Omega_0/H_\rho$, by the topographic $\beta$-effect, $2\Omega_0 (d\ln{h}/dz)$, where $h$ is the height of convective columns.


\section{Origin of the spatial asymmetry in the $y$-vorticity pattern} \label{appendix:origin_asymmetry}

\begin{figure}
\begin{center}
\includegraphics[width=0.9\linewidth]{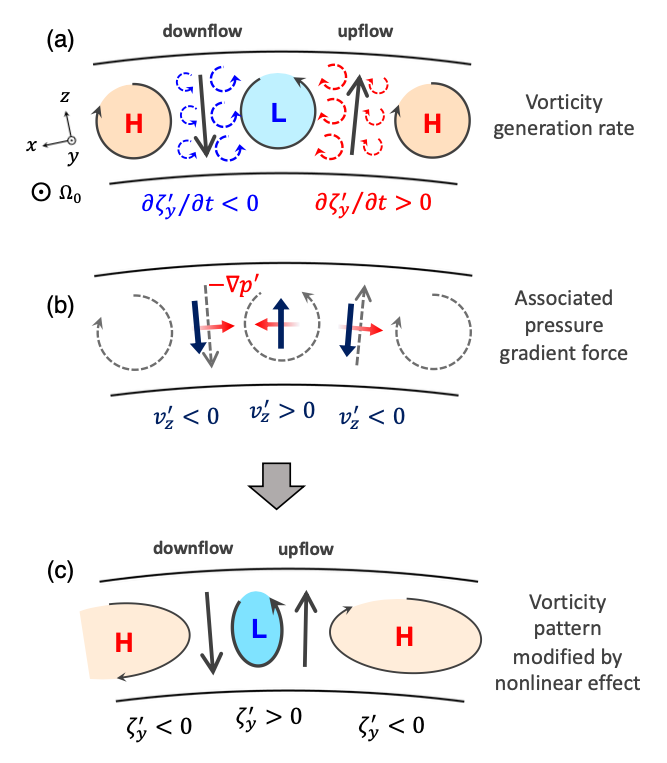}
\caption{
Schematic illustration explaining the origin of the asymmetry between counterclockwise and clockwise $y$-vortices.
(a) Vorticity generation rates in upflow and downflow regions.
Red and blue dashed circles with arrows indicate the generations of negative and positive $y$-vortices (corresponding to higher and lower pressure perturbations).
The vorticity generation is stronger (represented by larger circle size) inside the low-pressure column with $\zeta_y' >0$ due to the nonlinear effect.
(b) Directions of the pressure gradient forces (red arrows) and the flows required to balance these pressure gradient forces by their Coriolis forces (black arrows).
(c) Spatial pattern of the $y$-vortices modified by the nonlinear effect.
In all panels, the rotational axis points towards us.
}
\label{fig:asymmetry}
\end{center}
\end{figure}

A linear theory predicts that thermal Rossby modes at a fixed longitudinal wavenumber consist of positive and negative $y$-vorticity columns organized in an alternating pattern in longitude.
However, our simulations show significant asymmetry between the columns with $\zeta'_y >0$ and $\zeta'_y <0$:
That is, the columns with $\zeta'_y >0$ are stronger and more spatially localized in longitude compared to those with $\zeta'_y <0$ (see Figs.~\ref{fig:side_yave_MHD}a and b).
In this Appendix, we provide a physical explanation for the origin of this asymmetry based on a nonlinear effect.

We begin by examining the $y$-component of the vorticity equation
\begin{eqnarray}
 && \frac{\partial \zeta'_y}{\partial t} = [...] + \beta_{\rm nonlin} v'_z,
\end{eqnarray}
where the definition of the compressional $\beta$-effect is slightly modified to include the nonlinear advective term as
\begin{eqnarray}
    && \beta_{\rm nonlin} = - \frac{2\Omega_0 + \zeta'_y}{H_\rho}.
\end{eqnarray}
This means that the compressional $\beta$-effect is effectively enhanced in regions with $\zeta'_y >0$ and is suppressed where $\zeta'_y <0$.
Now, let us consider upflow and downflow regions located between positive and negative $y$-vorticity columns.
As already explained in Appendix~\ref{appendix:origin_tilt}, positive (negative) $y$-vorticity perturbations are generated in downflow (upflow) regions.
What is crucial here is that this vorticity generation becomes stronger (weaker) inside the positive (negative) $y$-vorticity columns, as schematically illustrated in Fig.~\ref{fig:asymmetry}a.
Since positive (negative) pressure anomalies are involved with negative (positive) $y$-vortices in a geostrophic fluid, additional pressure gradient forces are induced in a longitudinal direction.
As shown in Fig.~\ref{fig:asymmetry}b, these additional pressure gradient forces are directing retrograde in upflow/downflow regions and prograde inside the positive $y$-vorticity columns.
In order to balance these additional pressure gradient forces by the Coriolis force, vertical velocity perturbations are required.
Consequently, downflows (upflows) are enhanced (weakened) and upflow regions are shifted towards prograde direction by this nonlinear effect.
As depicted in Fig.~\ref{fig:asymmetry}c, these modifications lead to a more spatially localized structure for columns with $\zeta'_y >0$, while columns with $\zeta'_y <0$ become more spatially extended.


\end{appendix}

\end{document}